\documentclass[]{emulateapj}
\usepackage{natbib}
\usepackage{lscape}
\newcommand{\msun}{M_\odot}

\newcommand{\pc}{\rm{pc}}
\newcommand{\gyr}{\rm{Gyr}}

\newcommand{\tcl}{t_{\rm{cl}}}

\newcommand{\sing}{{\tt s-s}}
\newcommand{\bs}{{\tt b-s}}
\newcommand{\bb}{{\tt b-b}}
\newcommand{\mtb}{{\tt MTB}}
\newcommand{\se}{{\tt SE}}
\newcommand{\nbss}{N_{\rm{BSS}}}
\newcommand{\nbssc}{N_{\rm{BSS}, c}}
\newcommand{\nhb}{N_{\rm{HB}}}
\newcommand{\nhbc}{N_{\rm{HB}, c}}
\newcommand{\dens}{M_\odot \rm{pc}^{-3}}
\newcommand{\tage}{t_{\rm{age}}}

\shorttitle{Collisions and Blue Straggler Stars}
\shortauthors{Chatterjee et al.}

\begin{document}

\title{Stellar Collisions and Blue Straggler Stars in Dense Globular Clusters}

\author{Sourav\,Chatterjee}
\affil{Department of Astronomy, University of Florida, Gainesville, FL 32611.}
\email{s.chatterjee@astro.ufl.edu}
\author{Frederic A.~Rasio}
\affil{Center for Interdisciplinary Exploration and Research in Astrophysics (CIERA) and \\
	Dept. of Physics \& Astronomy, Northwestern University, 2145 Sheridan Rd, Evanston, IL 60208, USA. }
\author{Alison Sills}
\affil{Department of Physics and Astronomy, McMaster University, 1280 Main Street West, Hamilton, ON, L8S 4M1, CANADA. }
\and
\author{Evert Glebbeek}
\affil{Department of Astrophysics/IMAPP, Radboud University Nijmegen, PO Box 9010, 6500 GL, Nijmegen, The Netherlands. }


\begin{abstract}
 Blue straggler stars (BSS) are abundantly observed in all Galactic globular clusters (GGC) where 
data exist.  However, observations alone cannot reveal the relative importance of 
various formation channels or the typical formation times for this well studied population of 
anomalous stars.  Using a state-of-the-art H\'enon-type Monte Carlo code that includes  
all relevant physical processes, we create 128 models with properties typical 
of the observed GGCs.  These models include realistic numbers of single and binary 
stars, use observationally motivated initial 
conditions, and span large ranges in central density, concentration, binary 
fraction, and mass.  Their properties can 
be directly compared with those of observed GGCs.  
We can easily identify the BSSs in our models and determine their formation channels 
and birth times.  We find that for central densities above $\sim10^3\,\dens$ the dominant 
formation channel is stellar collisions while for lower density clusters, mass transfer in binaries 
provides a significant contribution (up to ~ 60\% in our models). 
The majority of these collisions are binary-mediated, 
occurring during 3-body and 4-body interactions.   
As a result a strong correlation between the specific frequency of 
BSSs and the binary fraction in a cluster can be seen in our models. 
We find that the number of BSSs in the core 
shows only a weak correlation with the collision rate estimator $\Gamma$ traditionally used by observers, 
in agreement with the latest Hubble Space Telescope (ACS) data. 
Using an idealized ``full mixing'' prescription for collision products, our models indicate that 
the BSSs observed today may have formed several Gyrs ago.  
However, denser clusters tend to have younger ($\sim 1\,\gyr$) BSSs.  
\end{abstract}

\keywords{methods: numerical --- methods: statistical --- blue stragglers --- stars: kinematics and dynamics --- globular clusters: general}

\section{Introduction}
\label{sec:intro}
Blue straggler stars (BSSs) are H-burning stars bluer and brighter (and therefore more massive) 
than the main-sequence 
turnoff \citep[of a star cluster;][]{1953AJ.....58...61S}.  It is impossible for normal single star evolution to create 
them.  Instead, BSSs must be created when a main-sequence (MS) star's mass is 
increased either via mass transfer in a binary (MTB) 
or by physical collision and merger with another star. Both processes can 
also increase the residual H-burning life of the star (rejuvenation) by bringing in new Hydrogen for burning in the core 
\citep[eg.,][]{1995ApJ...445L.117L,1996ApJ...468..797L,1997ApJ...487..290S,2001ApJ...548..323S,2002ApJ...568..939L,2009MNRAS.395.1822C} 
\footnote{Note that even if no new Hydrogen is supplied to the core, the star may still appear as a BSS during 
its own residual MS life if its present mass is higher than the MS turn-off.}.  
Since the BSSs are among the brighter 
members of a cluster, they are relatively easier to detect in photometric observations.  
All Galactic globular clusters (GGCs), where appropriate data exist, 
have shown this exotic stellar population \citep[e.g.,][]{2002yCat..33910945P,1995A&A...294...80F}.   
Many open clusters also show moderately large populations ($\sim 10$s) of BSSs 
\citep[e.g.,][]{2009Natur.462.1032M}; they are also found in some dwarf galaxies \citep[e.g.,][]{2007MNRAS.380.1127M} and 
even in the field \citep[e.g.,][]{2000AJ....120.1014P}. 

Stellar collisions can happen in many ways in a star cluster.  Two single stars can collide directly 
(s-s collision).  
If binaries are present, then binaries will interact with other single stars (b-s) or binaries (b-b).  
During b-s and b-b encounters, physical collisions can occur between two or more MS stars taking part 
in the interactions \citep{2004MNRAS.352....1F}.  

During MTB the details of the orbital evolution and mass transfer rate determine 
the final outcomes.  Mass transfer can be stable, such that a fraction of the companion mass is transferred 
to the MS star and the binary remains intact; or it can be unstable, leading to 
a more rapid dynamical evolution and the eventual merger of the two stars \citep{2009MNRAS.395.1822C}.  
Stellar mergers may also happen through interactions 
involving higher-order hierarchical systems.     
For example, a hierarchical {\it triple\/}, 
for certain configurations can increase the eccentricity of the inner orbit via the Kozai effect \citep{1962AJ.....67..591K}.   
If the eccentricity becomes sufficiently high, the two inner stars could then be driven to merge \citep{2009ApJ...697.1048P}.  
Although this may be a significant channel 
for BSS formation in less dense open clusters \citep{2009ApJ...697.1048P}, it is not clear 
whether these triples can survive long enough in denser GGCs to be 
effective.  

The interactions that produce BSSs depend 
on the cluster global properties, including the central density ($\rho_c$), binary fraction ($f_b$), distribution 
of initial binary orbital properties, and mass-segregation timescale.  
Thus the BSSs in a cluster can, in principle, provide a window onto the 
system's dynamical history.  In this respect the formation channel(s) of the BSSs and their 
typical formation times are of great interest.  Unfortunately, although observations in 
various wavelength bands have taught us a lot about the stellar properties of the 
BSSs, it is hard to determine how a particular BSS 
formed or how long ago the interaction occurred.  
Theoretical modeling is the only way to approach these questions.  

A popular approach so far to learn about BSS formation has been to carefully 
study the correlation between the number of BSSs (normalized in some way by another characteristic 
stellar population in the same region of the cluster) and some dynamically important cluster property, such as 
the collision rate estimator $\Gamma$ \citep{2006ApJ...646L.143P}, $f_b$ \citep{2008A&A...481..701S}, 
or the total cluster mass \citep[most recently][]{2013MNRAS.428..897L}.  
While these observed correlations can in principle teach us a lot about the possible 
formation channels of BSSs, direct understanding of these processes 
is impossible to achieve without careful modeling.  For example, a BSS created via MTB could actually 
be the result of some past dynamical encounter that changed the binary orbit 
in such a way that mass transfer occurred.  On the other hand, a 
collision may be mediated by resonant dynamical interactions involving binaries, blurring the
distinction between the simple ``collisional'' and ``binary'' origins often discussed in observational
studies \citep[e.g.,][]{2009Natur.457..288K,2011MNRAS.416.1410L}.  
In addition to these complications, it is also hard to determine observationally what the binary 
fraction in 
the cluster is at present, and it is impossible to know what it was in the past. Even in the complete absence
of binaries, the calculation of 
the $\Gamma$ parameter involves many properties of the cluster that are hard to measure accurately. 

The only way to explore and probe these complicated interactions in an evolving cluster is by 
numerical modeling of the full $N$-body system with stellar evolution, using a realistically large $N$, 
the full stellar mass function, and treating binary stars in detail. One must then also include 
all relevant physical processes, such as two-body relaxation, strong encounters 
including physical collisions, b-b and b-s interactions, MTB, tidal stripping in the Galactic potential, etc.  
In these models, if all interactions are recorded, in addition to the overall cluster 
properties and individual stellar properties at frequent enough output times, then all necessary 
information can be obtained about BSS formation and evolution. In particular, after BSSs have been identified 
in any snapshot of the system, the detailed past history of each BSS going all the way back to $t = 0$ 
can be extracted from the numerical data.       

For many years, this kind of realistic modeling of GGCs (particularly using high enough $N$ and $f_b$) 
remained very challenging because of the extreme computational  cost of direct $N$-body simulations.  
Hence, much more simplified models were used to study the BSS formation channels 
and other properties such as their radial distribution in the cluster \citep{2006MNRAS.373..361M}.  
Recently, modern H\'enon-type Monte Carlo codes \citep{1971Ap&SS..14..151H}, modified to include the relevant physical 
processes (in particular, strong interactions and single and binary stellar evolution), have opened new
possibilities 
\citep[e.g.,][]{2010ApJ...719..915C,2011MNRAS.410.2698G,2013MNRAS.429.1221H,2013MNRAS.429.2881C,2013MNRAS.431.2184G}. 

The current version of our Cluster Monte Carlo code (CMC) is well suited 
to the modeling of GGCs including all relevant physical processes, with large $N\sim10^5 - 10^6$ and realistic 
$f_b$ values. CMC has been developed and rigorously tested for over a decade 
\citep{2000ApJ...540..969J,2001ApJ...550..691J,2003ApJ...593..772F,2007ApJ...658.1047F,2010ApJ...719..915C,2012ApJ...750...31U,2013ApJS..204...15P,2013MNRAS.429.2881C}. Using this code it is now 
possible to produce large numbers of highly detailed cluster models covering the full range of observationally 
motivated initial parameters.  
For example, recent studies with CMC have explored a large range in parameter 
space and successfully created a library of cluster models that have properties 
very similar to the observed GGCs \citep{2010ApJ...719..915C,2013MNRAS.429.2881C}.  

Here we report results from our analysis of $128$ cluster models produced with CMC with finely sampled output
focusing on BSS formation and evolution.
These models include all relevant physical processes, and include realistic $N$, and $f_b$, and hence 
can be directly compared with observed 
GGCs without any need for rescaling.  In addition to all dynamical properties, single and binary stellar evolution 
is also followed in detail in these models.  
Thus our simulations provide an unprecedented opportunity to understand the dominant dynamical and stellar evolution
processes responsible for BSS production in clusters, spanning a large range in 
cluster parameters, and allow us to test for various correlations between BSSs and cluster properties. 

Previous works have pointed out that some dynamical properties such as 
core radius  ($r_c$) or $\rho_c$ for a cluster can 
have very different values according to different definitions used by theorists and observers 
\citep[most recently][]{2013MNRAS.429.2881C}. Hence, to avoid 
confusion, we restrict our analysis in this paper to the standard theoretical definitions used
in most $N$-body simulations \citep{1985ApJ...298...80C},
unless otherwise specified. A different and independent 
analysis of the same models is presented in \citet{2013arXiv1303.2667S}, where all relevant quantities are 
determined using the standard {\it observational\/} definitions. 
In this paper we focus on two basic questions: (1) What is the 
dominant physical process that creates the BSSs observed in GGCs? (2) With our adopted rejuvenation 
prescription \citep{2000MNRAS.315..543H,2002MNRAS.329..897H}, what is the typical age 
($t_{\rm{age}}$) of the observed BSSs? 
 
The paper is organized as follows.  
In Section\ \ref{sec:numerics} we describe briefly our simulation setup, explain how the 
BSSs are identified in the models, and provide definitions for various categories of BSSs based 
on their formation histories.  We also show comparisons between modeled and observed 
BSSs in terms of observable properties. We discuss the relative 
importance of the various formation channels in Section\ \ref{sec:branching}.  The time elapsed since 
formation for the BSSs in our models is explored in Section\ \ref{sec:dynamical_ages}.  
In Section\ \ref{sec:gamma} we investigate correlations of the number of BSSs in the cluster 
($\nbss$) with $f_b$ and $\Gamma$.    
We summarize and conclude in Section\ \ref{sec:conclude}.                
  
\section{Numerical Method}
\label{sec:numerics}
We use CMC and adopt a large grid of initial conditions over the range of 
values typical of the observed young massive clusters \citep[e.g.,][]{2007A&A...469..925S,2009gcgg.book..103S} to 
create $128$ detailed star-by-star models. 
All our simulated clusters have initial virial radii between $r_v=3$ and $4\,\pc$.  Initial $N$ is varied between 
$4$ and $10\times10^5$ stars.  The positions and velocities of the stars (and center of mass for the binaries) 
are assigned according to King profiles with 
$W_0$ between 
$4$ and $8$.     
We vary the initial $f_b$ between $0.05$ and $0.3$.
The masses of the single stars, or primaries in case of a binary, are chosen
from the IMF presented in \citet[][their Equations\ $1$ and $2$]{2001MNRAS.322..231K} 
in the stellar mass range $0.1-100\,\rm{M_\odot}$.  
Secondary binary companion masses
are sampled from a uniform distribution of mass ratios in the range $0.1\,\rm{M_\odot}-m_p$, where 
$m_p$ is the mass of the primary.  
The semi-major axes, $a$, for stellar binaries are chosen from a 
distribution flat in $\log a$ within physical limits, namely between
$5\times$ the physical contact of the components and the local 
hard-soft boundary \citep{2003gmbp.book.....H}.  Although initially all binaries in our models are hard at their respective positions, 
some of these hard binaries can become soft during the evolution of the cluster.  
The cluster contracts under two-body relaxation and 
the velocity dispersion increases making initially hard binaries soft.  Moreover, binaries 
sink to the core due to mass segregation where the velocity dispersion is higher than that at the initial binary positions.  
We include these soft binaries in our simulations until 
they are naturally disrupted via strong encounters in the cluster.  

The numerical setup and initial properties 
for these simulations are similar to the work presented in \citet{2013MNRAS.429.2881C} 
which shows that similar initial conditions result in models at the end of the simulation (at cluster age $\tcl \approx 12\,\gyr$) 
with properties including $r_c$, half-mass radius ($r_h$), $\rho_c$, total mass ($M$), and 
relaxation timescale at half-mass ($t_{rh}$) very similar to those observed in the GGCs.  
The differences in the initial conditions between these simulations and those presented in \citet{2013MNRAS.429.2881C} are 
as follows. Models with zero initial $f_b$ are discarded in this study since observations suggest that this is not the case 
for the GGCs in reality \citep{2008AJ....135.2155D, 2012A&A...540A..16M} \footnote{In fact, comparison between 
the observed BSS numbers in the GGCs and the numbers obtained via mock observations of these models indicates 
that even initial $f_b=0.05$ may be too low \citep{2013arXiv1303.2667S}. }.  
In addition, the models presented in this study explore 
a larger range in initial $f_b$, and $W_0$. 

While in most cases we present theoretical estimates of central quantities such as $r_c$ and $\rho_c$ 
\citep{1985ApJ...298...80C}, whenever we directly compare model quantities with observed GGC quantities, 
we use the corresponding ``mock observed'' values ($r_{c, \rm{obs}}$, and $\rho_{c,\rm{obs}}$) to remain consistent. 
Most recently \citet{2013MNRAS.429.2881C} discussed in detail various definitions for the cluster parameters and 
the relationships between theoretical 
and observed estimates of these quantities. Here, we estimate $r_{c,\rm{obs}}$ as the radius where the surface 
luminosity density drops by a factor of 2 from the central density \citep[e.g.,][]{1987degc.book.....S}. 
The values for $\rho_{c, \rm{obs}}$ are 
estimated from the peak central surface luminosity density and $r_{c, \rm{obs}}$ using the prescription given in 
\citet[][their Equation\ 4]{1993ASPC...50..373D} following \citet[][2010 edition]{1996AJ....112.1487H}. We assume 
$M/L = 2$ to convert luminosity to mass when required to determine the mock observed values. 
 
Our grid of simulations in this study covers star clusters at $t_{\rm{cl}} \approx 12\,\gyr$ with $r_c$ between 
$0.6$ and $2.8\,\pc$. This range in $r_c$ corresponds to $r_{c,\rm{obs}}$ in the range 
between $\approx 0.06$ and $5.07\,\pc$. Similarly, this grid spans clusters with $\rho_c$ 
between $\approx 7\times10^2$ and $10^5\,\dens$ which corresponds to $\rho_{c,\rm{obs}}$ between 
$\approx 3\times10^2$ and $4\times10^6\,\dens$. Final $M$ are between 
$1\times10^5$ and $4\times10^5\,\msun$, 
and corresponding $M_{\rm{obs}}$ values are similar. Among the $128$ 
models, $19$ have reached the binary-burning stage (thought to be equivalent to the post core-collapsed GGCs) 
before $\tcl = 12\,\gyr$, while the rest are still contracting 
\citep[i.e., they are non-core-collapsed GGCs,][]{2013MNRAS.429.2881C}.  
Relevant model properties are listed in Table\ \ref{tab:runlist}. For more detailed comparison between 
model and GGC properties
see \citet{2013MNRAS.429.2881C}. For mock observed BSS values 
for these models and further comparisons with GGC properties see \citet{2013arXiv1303.2667S}. 

\begin{figure*}
\begin{center}
\plotone{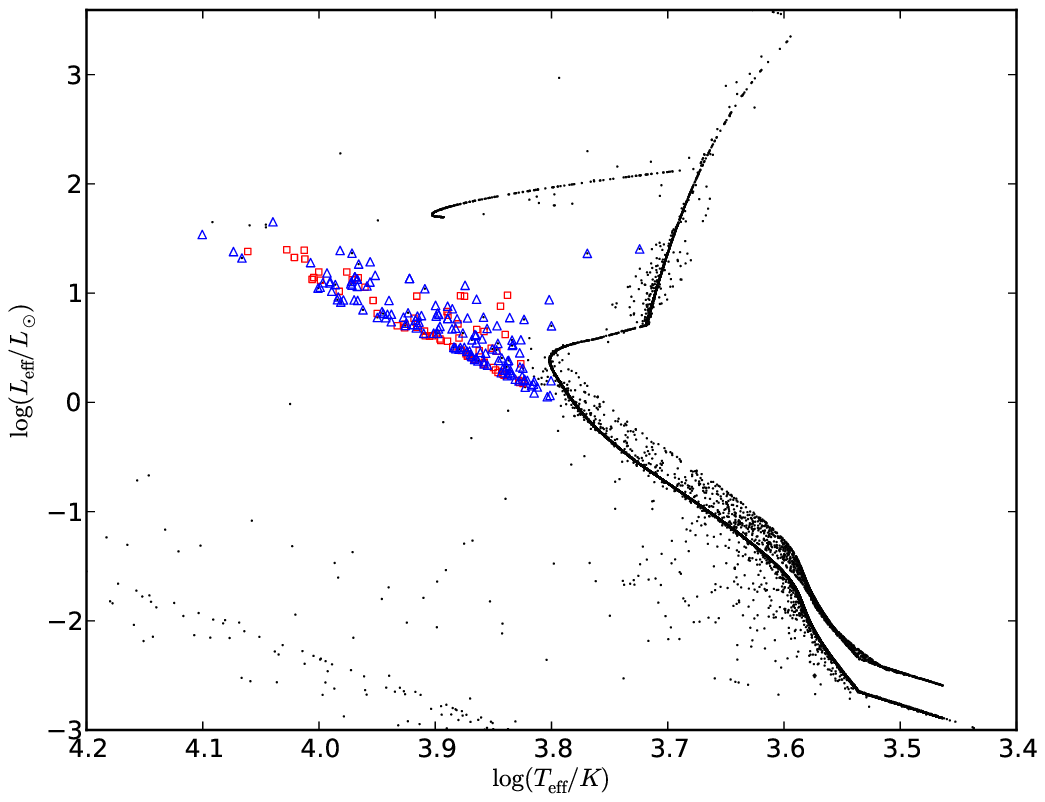}
\caption[HRD]{Example of a synthetic Hertzsprung-Russell (HRD) diagram for a representative 
model with many blue straggler stars ({\tt run110}, Table\ \ref{tab:runlist}). 
Dots denote all stars in the cluster. Red squares and blue triangles denote blue-straggler stars in single and binary 
systems, respectively at the integration stopping time of $\tcl=12\,\gyr$. 
Blue straggler stars are defined as H-burning stars with stellar masses $\geq 1.1\,M_{\rm{TO}}$, where 
$M_{\rm{TO}}$ is the main-sequence turn-off mass. BSSs in this model populate the expected region in the 
synthetic HRD. In addition, the single and binary main-sequences, giant branch, white-dwarf cooling sequence 
are clearly visible.     
}
\label{plot:hrd}
\end{center}
\end{figure*} 
\subsection{Definition of blue straggler stars in our models}
\label{sec:bss_def}
We define BSSs as stars that are still in the H-burning stage but with 
masses $>1.1\,M_{\rm{TO}}$, where, $M_{\rm{TO}}$ is the MS turn-off mass of the cluster.  
Note that in real clusters BSSs are chosen from a color-magnitude diagram by a defining box for 
the BSSs that depend on, for example, the quality of the data, the filters used for the observation, 
and the scatter width of the MS \citep[e.g.,][]{2007ApJ...661..210L,2013arXiv1303.2667S}.  
The mass-based criterion is more theoretically motivated and is much simpler to use for all models (although, inaccessible to observers).  
We find that in our models the BSSs identified using the mass-based criterion 
populate the expected region of synthetic Hertzsprung-Russell diagram (HRD) 
created using our models (Figure\ \ref{plot:hrd} for an example).  
Observationally motivated selection boxes are 
adopted to extract BSSs from these same models in \citet{2013arXiv1303.2667S} and we find that the 
numbers obtained from either criteria are tightly correlated. Furthermore, the selected BSSs using 
either criteria do not show any systematic differences in their formation channels, formation ages, or radial 
positions. Since in this study we restrict ourselves to the theoretical understanding of the BSSs, in particular 
to understand the dominant formation channels and typical ages since formation, we simply adopt the 
much simpler theoretically motivated mass-based definition for BSSs for this study.    

\subsection{Classification of blue-straggler stars based on their formation channels}
\label{sec:classification}
At the last snapshot of the model ($\tcl\sim12\,\gyr$) we identify the BSSs following the criterion described in Section\ \ref{sec:bss_def}.  
The full dynamical history 
for each of these BSSs is retrieved.  
We identify the interaction (such as s-s, b-s, or b-b collisions, 
or MTB) that had increased the mass of this star for the {\em first time} such that it would be 
identified as a BSS at $\tcl$.  
We classify each BSS into one of $5$ channels according to this first interaction.  
The $5$ channels are: stable mass transfer in a binary denoted as 
\mtb, binary-stellar-evolution-driven merger denoted as \se, single-single collisions denoted as 
\sing, collisions mediated by binary-single interactions denoted as \bs, and collisions mediated by 
binary-binary interactions denoted as \bb.  Note that some BSSs may have a complicated 
history.  For example, after a mass transfer episode in a binary, a MS star may have attained 
a BSS mass;  later on this star could still collide with another MS star, 
further growing in mass and getting rejuvenated for a second time.  For BSSs that are formed via 
such multiple channels, {\it only the first one} is considered for the classification to avoid ambiguity.  
In this particular example the formation channel for this BSS would be classified as \mtb.  
Nevertheless, such complicated multiple episodes of rejuvenation of the same MS 
star is not frequent.  There is another source of complication.  A BSS classified as \mtb\ 
may have had interacted dynamically prior to the Roche-lobe overflow that had not changed its mass.  
Such an interaction would change the orbital properties of the binary, sending it along a very different 
evolutionary pathway that led to the mass transfer event that ultimately created the BSS.  
These perturbed \mtb\ or \se\ channels are duly noted and discussed.  However, interactions of this type 
do not change the formation classification.  

\subsection{Comparison of Model Blue Stragglers with Observed Blue Straggler Properties}
\label{sec:comparison}
Here we briefly investigate whether the modeled BSS properties are consistent with some of the 
well studied observed BSS properties in the GGCs. Note that these are motivated to add further 
validation of the model BSSs. More detailed analysis of these properties are potentially interesting 
but beyond the scope of this study and we differ that to \citet{2013arXiv1303.2667S} and future work. 
\subsubsection{Radial Distribution}
\label{sec:rdist}
One of the observationally accessible properties of the BSSs in the GGCs is their radial distribution.  
The radial distribution of the BSSs in a cluster depends on a collection of the host cluster's properties 
including $\rho_c$, relaxation time, and $f_b$.  Thus analysis of the radial distribution of the 
BSSs has been proposed as a tool to observationally constrain the dominant formation channel, and 
various dynamical properties of the host GGCs.  As a result, the radial distribution of the BSSs 
has been studied with interest for many GGCs via observations 
\citep[e.g.,][]{1992AJ....104.1831F,2008ApJ...681..311D,1995A&A...294...80F,2008ApJ...679..712B,2008ApJ...677.1069D,2008yCat..34830183M,2008A&A...483..183M,2012Natur.492..393F} as well as via theoretical 
modeling \citep[e.g.,][]{2006MNRAS.373..361M}.  Three qualitatively different types of radial distributions 
are observed among the GGCs, namely, single and bimodal, and flat distributions 
\citep[for a compilation see ][]{2012Natur.492..393F}. We check whether the radial distributions of our 
model BSSs are consistent with those observed in the GGCs.  

\begin{figure*}
\begin{center}
\plotone{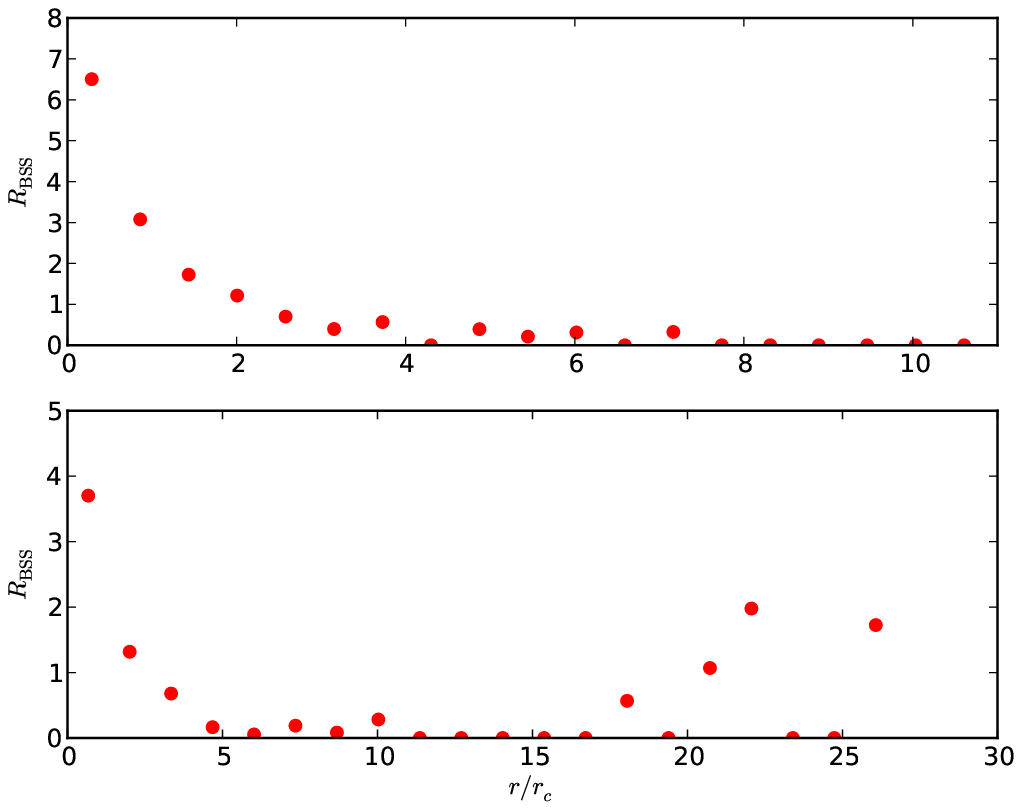}
\caption[radial distribution]{Examples of two qualitatively different radial distributions for our model BSSs.  
Vertical and horizontal axes denote the double-normalized ratio $R_{\rm{BSS}}$ 
(Equation\ \ref{eq:rbss}) and radial position in units of $r_c$, respectively.  
Top panel shows an example of centrally peaked radial distribution ({\tt run108}; Table\ \ref{tab:runlist}).  
Bottom panel shows an example of a bimodal distribution ({\tt run110}), one peak near the center $r/r_c < 5$ and 
a second peak at large $r/r_c \approx 22$.  A zone of avoidance is observed at $r/r_c$ between about 
$5$ and $17$.  }
\label{plot:radial_dist}
\end{center}
\end{figure*} 
To find the relative frequency of the BSSs in the cluster as a function of their radial positions we 
define a double-normalized ratio $R_{\rm{BSS}}$ following \citet{2012Natur.492..393F} given by 
\begin{center}
\begin{equation}
\label{eq:rbss}
R_{\rm{BSS}} = \frac{ \delta N_{\rm{BSS}}/N_{\rm{BSS}} } { \delta L/L_{\rm{cl}} }, 
\end{equation}
\end{center}
where $\delta N_{\rm{BSS}}$ denotes the number of BSSs within some radial bin, $N_{\rm{BSS}}$ 
denotes the total number of BSSs, $\delta L$ denotes the stellar luminosity within that bin, and $L_{\rm{cl}}$ denotes 
the total luminosity of the cluster.  $R_{\rm{BSS}}$ is a measure of whether the BSSs follow the luminosity density 
profile of the cluster or they are over/under abundant in some parts.  

Figure\ \ref{plot:radial_dist} shows $R_{\rm{BSS}}$ as a function of the 
radial position in units of the core radius.  
In our models we find examples of bimodal distributions (bottom panel, observed, e.g., in M55) as well 
as centrally peaked distributions \citep[top panel, observed, e.g., in M80,][and references therein]{2012Natur.492..393F}.  
In our models we do not find flat radial distributions 
observed, for example, in NGC 2419 \citep{2008ApJ...681..311D}.  This is expected since all our models 
are relaxed and hence a non-segregated radial distribution for the BSSs is not expected in our models.       
\citet{2012Natur.492..393F} suggest that the minimum distance $r_{\rm{min}}$ from the center where 
the central peak of $R_{\rm{BSS}}$ ends in units of $r_c$ is anti-correlated with the core relaxation timescale 
due to mass-segregation effects of the relatively high-mass BSSs.  Results from our models are 
consistent with this finding.  However, we find that the determination of $r_{\rm{min}}$ can 
have large errors and depends strongly on the radial bins used to calculate $R_{\rm{BSS}}$.  

\subsubsection{Core Blue Straggler Number and Core Radius}
\label{sec:rc_vs_nbssc}
\begin{figure*}
\begin{center}
\plotone{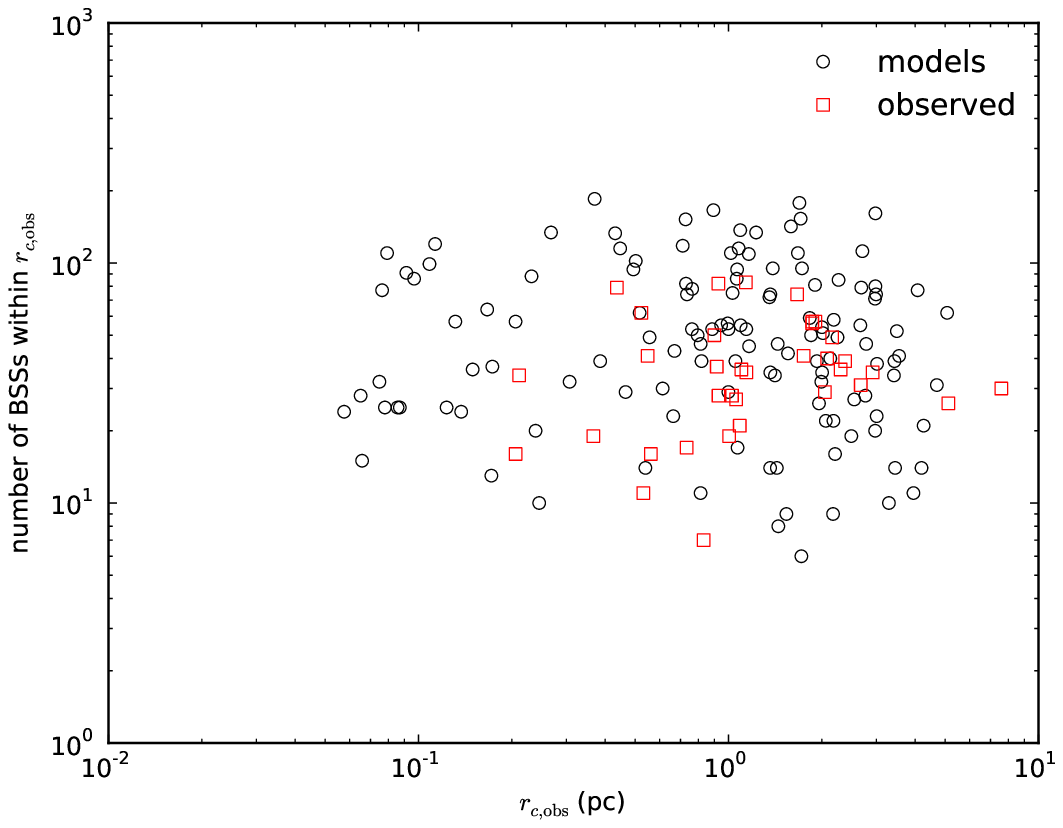}
\caption[$N_{BSS}$ vs $r_c$]{Number of BSSs in the core ($N_{\rm{BSS},c}$) vs $r_{c,\rm{obs}}$.  
Black circles and red squares denote model and observed BSSs, respectively.  
The observed $\nbssc$ values are obtained from the HST/ACS survey 
\citep{2007AJ....133.1658S} as selected by \citet{2011MNRAS.415.3771L}.  
Our models populate very similar $r_{c,\rm{obs}}$ values compared to the observed GGCs. However, 
our models show slightly higher $N_{BSS,c}$ numbers due to the mass-based selection criteria. 
If these models could be observed using the HST/ACS filters and the BSSs were identified using a 
more observationally motivated selection box $\nbssc$ would decrease by about a 
factor of $2$ \citep{2013arXiv1303.2667S}. 
}
\label{plot:nbscvsrc}
\end{center}
\end{figure*} 
Are the number of BSSs in the core ($\nbssc$) in our models consistent with those in the GGCs? 
Figure\ \ref{plot:nbscvsrc} shows $\nbssc$ as a function of $r_{c,\rm{obs}}$ both 
for our models and for observed GGCs. Note that since we are directly comparing with observed BSSs, we use 
$r_{c, \rm{obs}}$ for this comparison. The BSSs are, however, chosen using the same mass-based criteria. The 
observed BSS counts are obtained from the ACS survey data \citep{2007AJ....133.1658S} as selected by 
\citet{2011MNRAS.415.3771L}. 
Our models clearly have very similar core radii compared to the observed GGCs. We identify 
a slightly larger $\nbssc$. More observationally motivated selection criteria 
presented in \citet{2013arXiv1303.2667S} for the same models show that if these systems 
were observed, the number of BSSs actually detected would have been smaller by about a factor of 2. Thus 
we conclude that $\nbssc$ as produced in our models and the core radii both are consistent with 
the observed Milky Way populations.  

In addition to these caveats about different identification criteria for BSSs, one should 
also note that, within BSE, and the current version of CMC, the rejuvenation prescription used for all mergers, 
whether from collisions or binary evolution, and MTB assumes ``full mixing" 
\citep[for details see][]{2000MNRAS.315..543H,2002MNRAS.329..897H}. 
This is an oversimplification.  In reality, the amount of Hydrogen core mixing and thus the degree of 
rejuvenation of the MS star depends on the kinematic details of the encounter 
\citep[e.g.,][]{1997ApJ...487..290S,2001ApJ...548..323S}.  
Because of this simplification the lifetimes of the BSSs after rejuvenation in our models are 
actually {\em upper limits\/} (at least for non-rotating stars).  
This may also lead to an over-production of BSSs in our 
models compared to real clusters. 
\section{Relative Importance of Blue Straggler Formation Channels}
\label{sec:branching}
\begin{figure*}
\begin{center}
\plotone{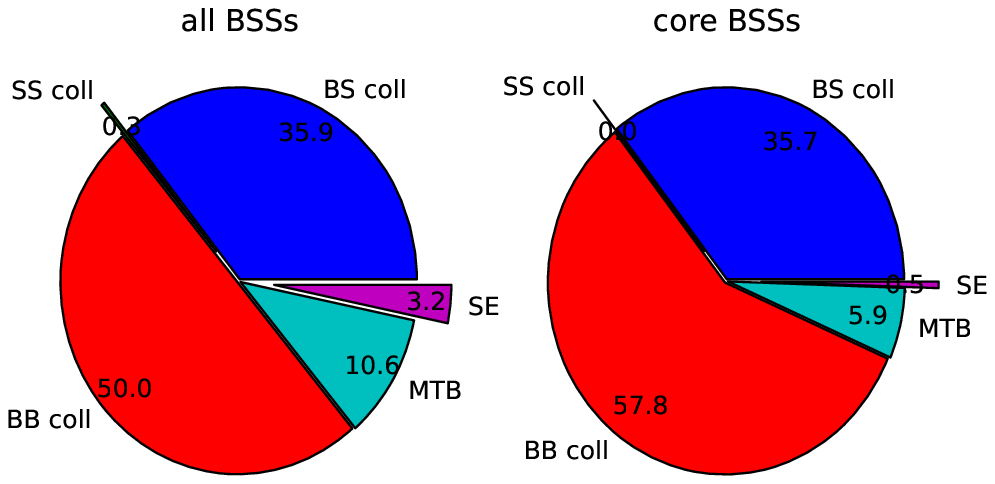}
\caption[BSS branching ratios]{Example of branching ratios for the various formation channels of 
BSSs at $\tcl = 12\,\rm{Gyr}$ ({\tt run32}; Table\ \ref{tab:runlist}). This model has a moderate central density 
($\rho_c \approx 7\times10^3\,\dens$; Table\ \ref{tab:runlist}). This particular 
model was chosen as an example due to its high number ($312$) of BSSs.  
The dominant formation channel is stellar collisions in this model. Among the collisional 
BSS production channels collisions mediated by \bb\ interactions dominate followed by 
\bs\ mediated collisions. \sing\ collisions are rare. \mtb\ is the dominant 
channel among the binary-stellar-evolution driven BSS production channels. \se\ 
contribute only to a small degree.     
}
\label{plot:pie}
\end{center}
\end{figure*} 
In this section we investigate the relative contributions from various BSS formation channels.  
Figure\ \ref{plot:pie} shows an example of branching ratios for the BSSs in the whole 
cluster, and in the core, for model {\tt run32} (Table\ \ref{tab:runlist}).  This is a model 
with moderate central density $\rho_c = 7\times10^3\,\rm{\msun pc^{-3}}$.  
Collisions produce most of the BSSs in this model.  Among the collisional BSSs, 
the \sing\ channel contributes the least.  The binary mediated collisional channels, namely 
\bs\ and \bb\ contribute about $36\%$ and $50\%$, respectively.  BSSs produced via the \mtb\ 
channel contributes about $11\%$ and \se\ contributes only $3.2\%$.  As expected collisional channels 
dominate even more among the core BSSs.  
Interestingly, although the \bs\ and \bb\ contributions increase if only core BSSs are considered 
instead of all BSSs in the cluster, \sing\ contribution decreases.  
This trend is seen often in 
our models because of the low number density of single MS stars within the core at late times, 
due to mass segregation.  

About $60\%$ of all BSSs formed via \mtb\ or \se\ in this example model have had at least one prior strong 
encounter. 
Thus for this particular model a significant 
fraction of even the \mtb\ and \se\ BSSs are not created via pure binary stellar evolution, and 
have been dynamically modified before binary stellar evolution could 
rejuvenate them.  

\begin{figure*}
\begin{center}
\plotone{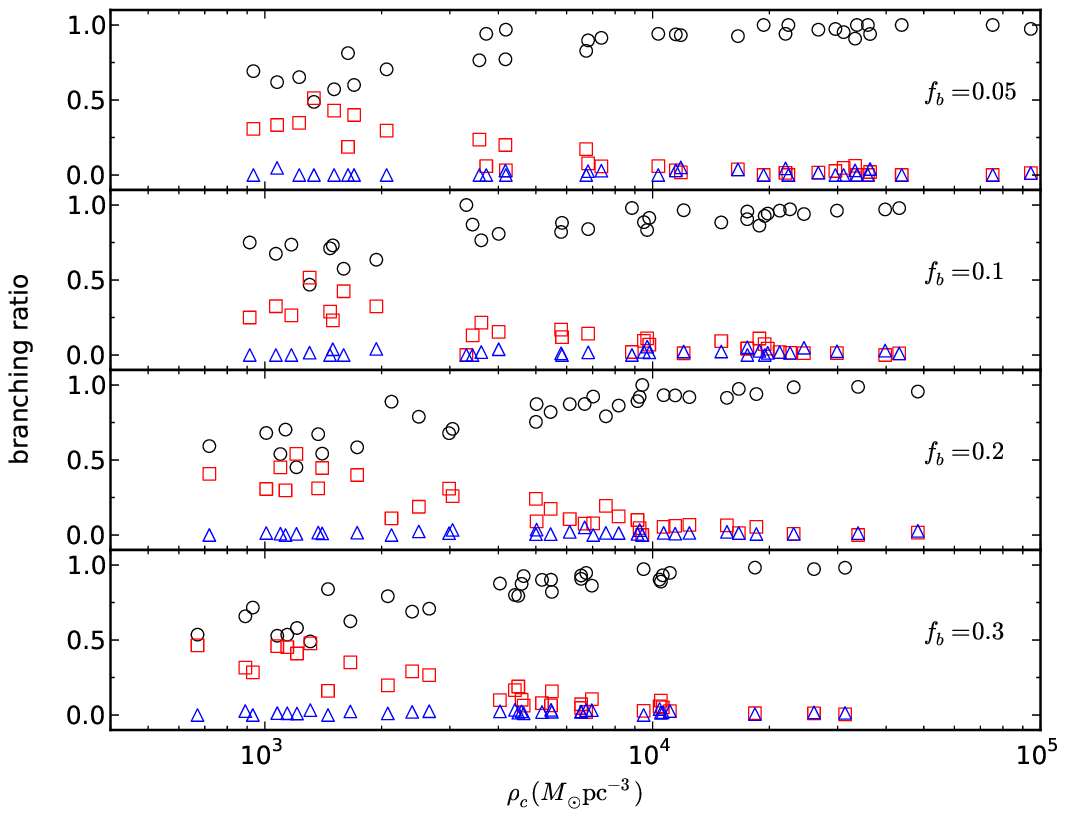}
\caption{Branching ratios of various formation channels for all BSSs in our models 
vs the respective cluster's $\rho_c$. The different panels from top to bottom show subsets of 
models with different initial $f_b$ in an increasing order. The respective initial $f_b$ values are 
shown in each panel. The circles, squares, and triangles 
denote collisional, \mtb, and \se\ channels, respectively.  Clearly, for $\rho_c >$ few 
$\times 10^3\,\rm{\msun pc^{-3}}$ collisional formations channels dominate for all initial $f_b$.  BSSs from \se\ are rare 
throughout the full range in $\rho_c$ and $f_b$ in our models. 
}
\label{plot:rhoc_branching}
\end{center}
\end{figure*} 
\begin{figure*}
\begin{center}
\plotone{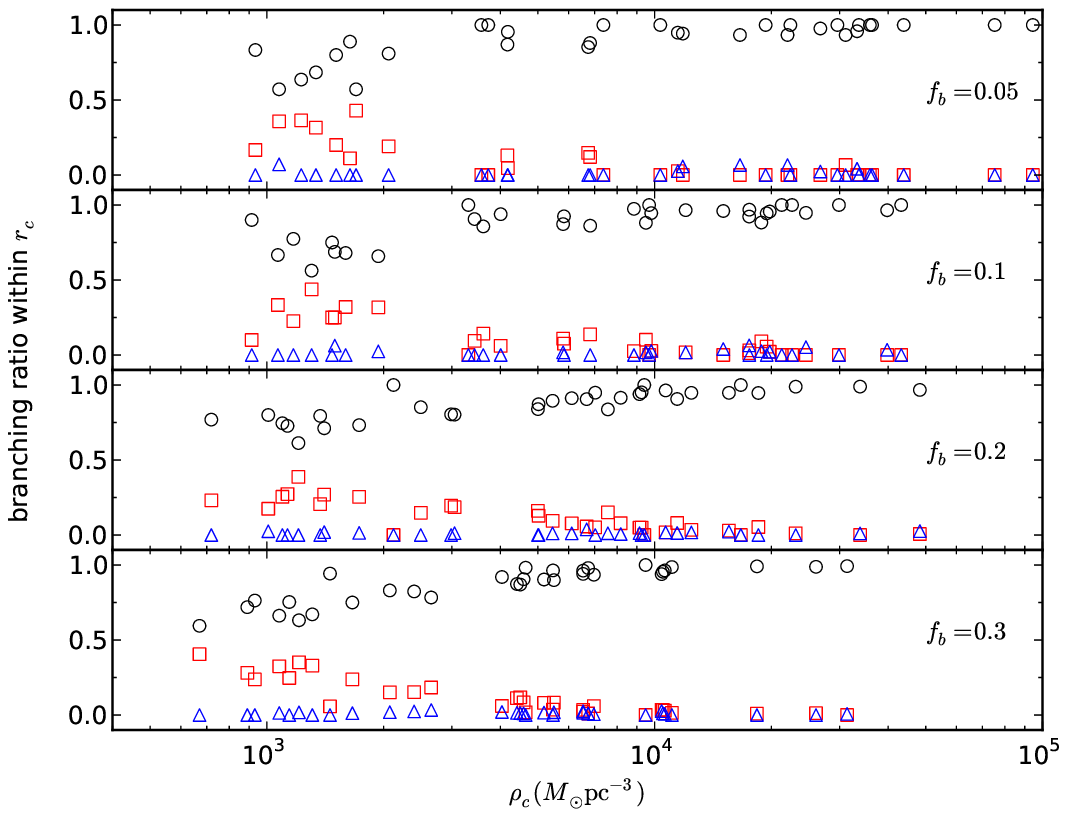}
\caption{Same as Figure\ \ref{plot:rhoc_branching} but including only the BSSs within the core.  
Collisional channels dominate the core BSSs throughout the full range in $\rho_c$ and $f_b$.  
\se\ mergers are rare throughout the range 
of $\rho_c$ in our models.  \mtb\ channel only becomes comparable to collisional channels 
if $\rho_c < 10^3\,\rm{\msun pc^{-3}}$.    
}
\label{plot:rhoc_branching_core}
\end{center}
\end{figure*} 

Branching ratios for various BSS production channels are expected to depend on 
$\rho_c$ and $f_b$ in a cluster. We divide our full set of models 
into subsets based on the initial $f_b$ and present the branching ratios for each subset 
separately as a function of $\rho_c$.  

Figure\ \ref{plot:rhoc_branching} shows the branching ratios for 
\sing, \bs, \bb, \mtb, and \se\ formation channels as a function of $\rho_c$. The different panels 
show subsets of models with different initial $f_b$.
For models with $\rho_c >$ few $\times 10^3\,\rm{\msun pc^{-3}}$ collisional channels 
clearly dominate.  In lower density clusters contribution from \mtb\ increases.  Throughout the 
full range in $f_b$ in our models ($f_b = 0.05$ -- $0.30$), the same trend is observed.  
This indicates, that even if $f_b$ is increased (within the studied range in $\rho_c$) 
the relative contribution from \mtb\ does not increase proportionally.  
While increasing $f_b$ increases the probability that a BSS can 
be created via \mtb\ (simply because of the increased number of binaries), 
the increased number of binaries also increases the \bs\ and \bb\ BSS formation 
rates.  Typical $\rho_c$ values in the GGCs are orders of magnitude higher than $\sim 10^3\,\rm{\msun pc^{-3}}$ 
\citep[e.g.,][]{1996AJ....112.1487H,2010ApJ...719..915C,2013MNRAS.429.2881C}.  Hence, we conclude 
that the collisional channels dominate production for the BSS populations observed in the majority of the GGCs.  
If only the core BSSs are considered, 
the importance of the collisional channels increases as expected.  
We find that throughout the full range of $\rho_c$ in our models collisional channels 
dominate the formation of core BSSs (Figure\ \ref{plot:rhoc_branching_core}).  

\begin{figure*}
\begin{center}
\plotone{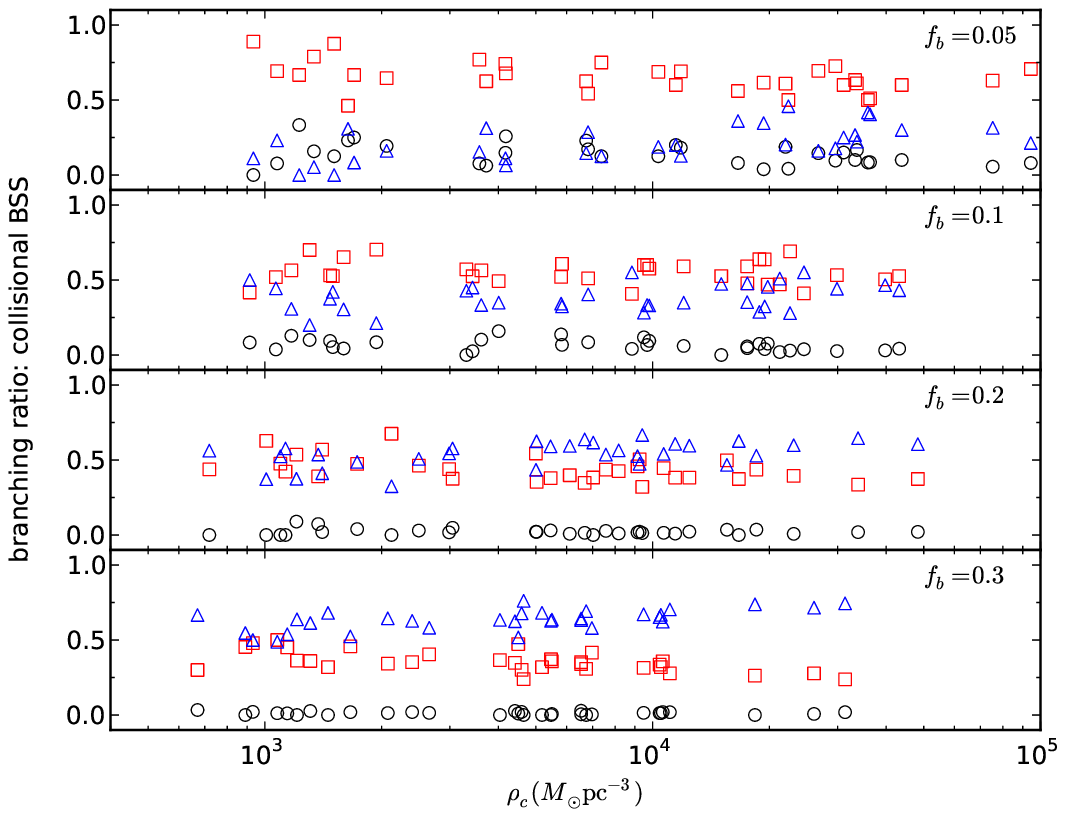}
\caption{Branching ratios between various collisional BSS formation channels vs $\rho_c$.  Squares, triangles, 
and circles denote \bs, \bb, and \sing\ collisions, respectively.  Each panel shows a subset of models 
with a given value of initial $f_b$ as mentioned on each panel.  
Throughout the studied range of $\rho_c$ and $f_b$ 
contribution from \sing\ channel is minor. Collisions from \bs\ and \bb\ dominate. For high 
$f_b = 0.3$ values \bb\ collisions dominate over \bs\ collisions in creation of BSSs.    
}
\label{plot:rhoc_branching_coll}
\end{center}
\end{figure*} 
\begin{figure*}
\begin{center}
\plotone{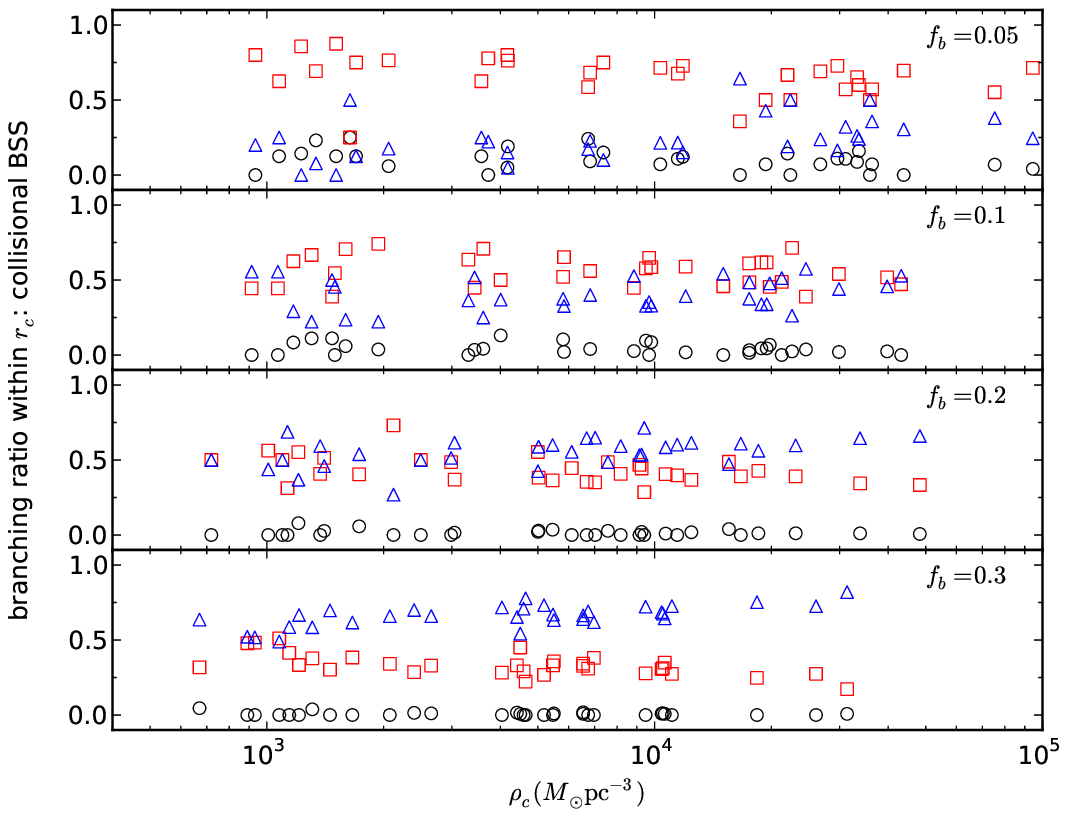}
\caption{Same as Figure\ \ref{plot:rhoc_branching_coll} but including only the BSSs within the core.  
}
\label{plot:rhoc_branching_coll_core}
\end{center}
\end{figure*} 
Although binary stellar evolution is not the dominant channel for a large range of central densities 
typical of GGCs, binaries themselves are nevertheless crucial for significant BSS production.  
Figure\ \ref{plot:rhoc_branching_coll} and \ref{plot:rhoc_branching_coll_core} show the relative contributions from various collisional 
formation channels for the BSSs in the whole cluster and in the core, respectively, 
as a function of $\rho_c$.  For the full range in $f_b$ and $\rho_c$ in our models 
BSSs produced via \sing\ are rare.  Binary-mediated collisions (\bs\ or \bb) dominate.  
As $f_b$ increases, so does the relative contribution from \bb.  For example, for models with initial $f_b = 0.05$, 
the \bs\ channel contributes more to the collisional BSSs compared to the \bb\ 
channel for all $\rho_c$ values.  In contrast, for initial $f_b = 0.3$ the contribution from the \bb\ channel 
dominates for all $\rho_c$ values we have considered.  

\begin{figure*}
\begin{center}
\plotone{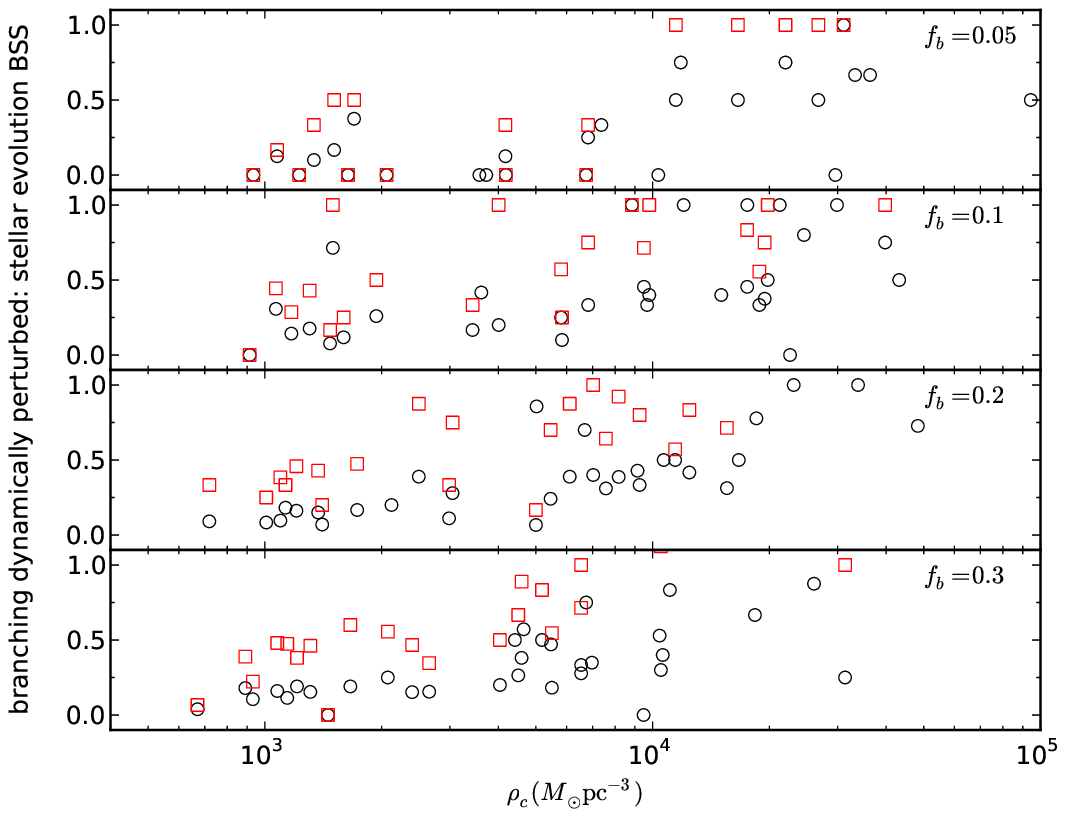}
\caption{
Fraction of BSSs created via \mtb\ and \se\ mergers that had been dynamically perturbed before 
the change in mass, with respect to all BSSs formed via \mtb\ and \se\ channels, as a function of $\rho_c$. 
Only perturbations before the production of the BSSs are taken into account. These BSSs would have not 
been produced with the same properties had there not been a previous encounter. Once produced, these 
BSSs may interact further; these fractions therefore indicate a lower limit to the fraction of BSSs created 
via binary stellar evolution and are dynamically altered. Black circles and red squares denote all BSSs in 
the models and only those within the core, respectively. Each panel from top to bottom show models with 
initial fb= 0.05, 0.1, 0.2, and 0.3, respectively.
}
\label{plot:rhoc_branching_mtb}
\end{center}
\end{figure*} 
Figure\ \ref{plot:rhoc_branching_mtb} shows the fraction of BSSs produced via binary stellar evolution 
(\mtb\ and \se\ collectively) that have been dynamically perturbed before the mass transfer 
process has occurred (with respect to all BSSs produced via \mtb\ and \se). 
These interactions can potentially alter the orbital properties of these binaries initiating 
very different evolutionary pathways for them.  BSSs formed from these disturbed binaries clearly 
would not have formed the BSSs with the exact same properties and at the same times without these encounters.  
As expected, for the models with higher $\rho_c$ values \mtb\ and \se\ BSSs have a higher chance of being 
perturbed before production.  For a given $\rho_c$ a range of these values are also seen.  This is due to the 
different concentrations for these clusters.  Models with initially lower concentrations  produce a smaller fraction of perturbed 
\mtb\ and \se\ BSSs.  The majority of the models with $\rho_c \gtrsim 10^4\,\dens$ produce more perturbed 
\mtb\ and \se\ BSSs than unperturbed.  Low density ($\rho_c \lesssim 10^3$) models on the other hand will have 
a relatively larger fraction of BSSs via the \mtb\ and \se\ channels created from the primordial undisturbed binaries. 
Similar results were also found previously by \citet{2008A&A...481..701S}. Our results support their conclusion 
that for $\rho_c\lesssim10^3\,\dens$ stellar evolution channels contribute $\gtrsim50\%$ of the BSSs, and a high 
fraction of them are produced from dynamically unmodified binaries.      

\section{Time since formation for the blue straggler stars}
\label{sec:dynamical_ages}
One of the biggest uncertainties in any star-by-star $N$-body model using analytical formulae 
\citep{2000MNRAS.315..543H} for hydrodynamic mergers and stellar evolution is the degree of rejuvenation. 
Unfortunately, the degree of rejuvenation depends on the detailed kinematics of the 
encounter \citep[e.g.,][]{1997ApJ...487..290S,2001ApJ...548..323S} as well as the stellar properties such 
as rotation, and the ages of the parent stars 
\citep{2009MNRAS.395.1822C}. Incorporating these details is beyond the scope of this study and can 
only be done via live stellar and binary evolution codes that can 
rigorously incorporate these physical effects while the whole cluster is dynamically evolving. This is a hard task 
and has yet to be achieved fully self-consistently by any group. Hence, although the analytical formulae 
and prescriptions used in BSE \citep{2000MNRAS.315..543H,2002MNRAS.329..897H} are known 
to over-estimate the degree of rejuvenation \citep[e.g.,][]{2008A&A...488.1017G}, for example, for collision 
products of non-rotating stars, this remains currently the state of the art. 
For the purpose of this study, however, we expect that 
the {\it relative contributions\/} from 
\mtb, \se, and various collisional channels should not change by much given that all channels are 
treated using maximal mixing within BSE, hence rejuvenation may be similarly over-estimated in 
all cases. 

Despite these caveats, 
it is important to study the distributions of $t_{\rm{age}}$ for our model BSSs since 
they provide us with useful upper limits. The typical distributions of $t_{\rm{age}}$ are also interesting in light of the current 
understanding that most GGCs (those observationally categorized as non-core-collapsed) are still contracting 
\citep[e.g.,][]{2013MNRAS.429.2881C}. As a result, relevant global cluster 
properties such as $\rho_c$, $r_c$, and $r_h$ have not attained steady-state values typical of 
core-collapsed, ``binary-burning'' clusters \citep{2013MNRAS.429.2881C}. Hence, for these contracting 
clusters it is not sufficient to know the present-day observed cluster properties to fully understand the formation efficiency 
of BSSs. Instead, the past dynamical history over at least the 
last $\sim t_{\rm{age}}$ should be taken into account. 
Of course the results (based on the simple full mixing assumption) presented here can also be directly compared 
with 
future results from simulations incorporating more sophisticated rejuvenation treatments 
when they become available.    

\begin{figure*}
\begin{center}
\plotone{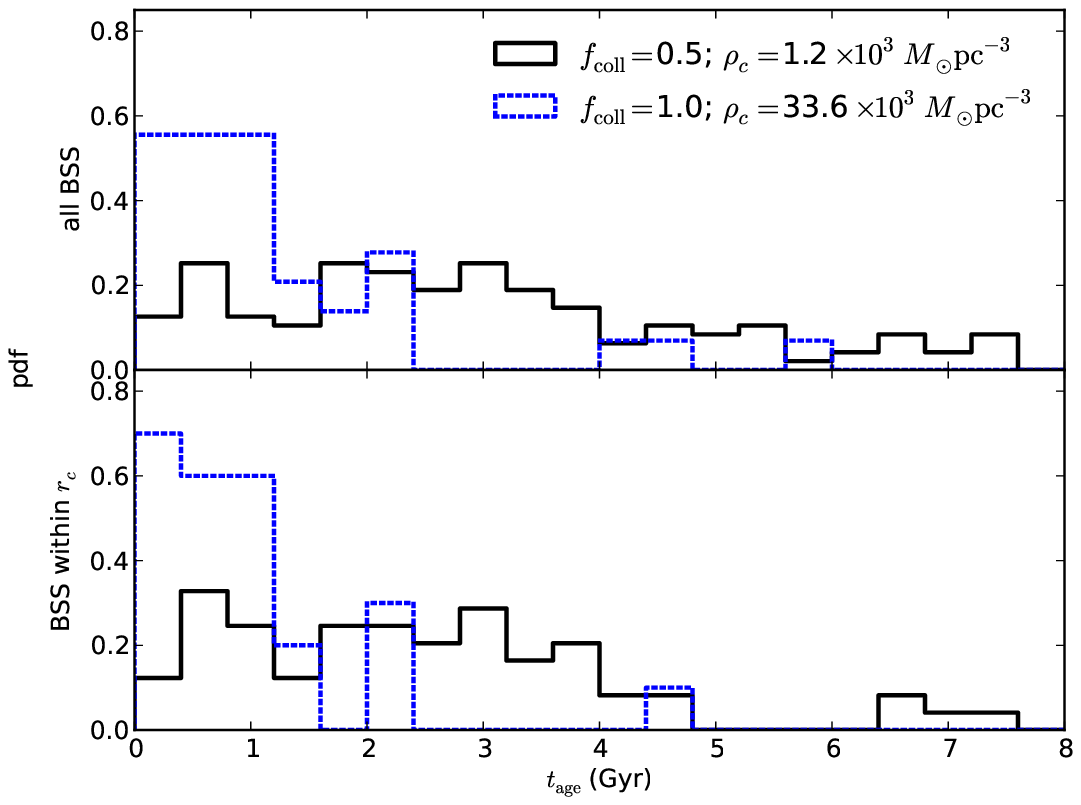}
\caption{Examples for distributions of the time elapsed since formation ($t_{\rm{age}}$) 
for the BSSs for two models, one for low ({\tt run60}) 
and another for high $\rho_c$ ({\tt run17}; Table\ \ref{tab:runlist}) 
with central densities varying by over an order of magnitude. The models were chosen such that each 
has $\nbss>21$ and have extreme contributions from collisional channels (0.5 for {\tt run60} and 1 for {\tt run17}). 
The solid (black) and dashed (blue) lines represent 
the low and high density clusters, respectively.  
The top and bottom panels are 
for all BSSs and those within $r_c$, respectively.  
For \mtb\ BSSs the formation time $t_{\rm{form}}$ is set at the beginning of Roche-lobe overflow.    
The BSSs in the low-density model shows a 
larger age-spread compared to those in the high-density model.  Low-density models have a larger relative 
contribution from \mtb\ channel.  Since \mtb\ BSSs are typically less massive compared to 
those created via collisions, the residual lifetime for the \mtb\ BSSs is longer than those 
produced via collisions.  
}
\label{plot:example_age}
\end{center}
\end{figure*} 
Figure\ \ref{plot:example_age} shows examples of 
the distributions of $\tage$ for two of our models. 
These models are chosen such that at least $21$ BSSs are created (to ensure enough statistics) 
and the relative contributions from collisional channels $f_{\rm{coll}}$ have extreme values 0.5 ({\tt run60}), 
and 1 ({\tt run17}; Table\ \ref{tab:runlist}). The central densities differ by a little over an order of magnitude.  

The BSSs in the high-$f_{\rm{coll}}$ (hence high-$\rho_c$) model are typically younger compared to those in the 
low-$f_{\rm{coll}}$ model.   
The broadly peaked distributions seen here are qualitatively similar in all our models.  
However, the exact positions of these peaks depend on the cluster properties.    
For the low-density model the bulk of the BSSs have ages between $\tage \approx 0$ and $7\,\rm{Gyr}$.    
In contrast, the BSSs in the high-density model typically have $\tage$ between $0$ and $3\,\gyr$.  
This is consistent with 
previous results suggested for the observed BSS population in 47 Tuc \citep{2000ApJ...535..298S}.  
Half of all BSSs in the low-density model is contained within an age 
of $\approx3\,\gyr$, whereas, for the high-density model this value is $\approx0.9\,\gyr$.           

\begin{figure*}
\begin{center}
\plotone{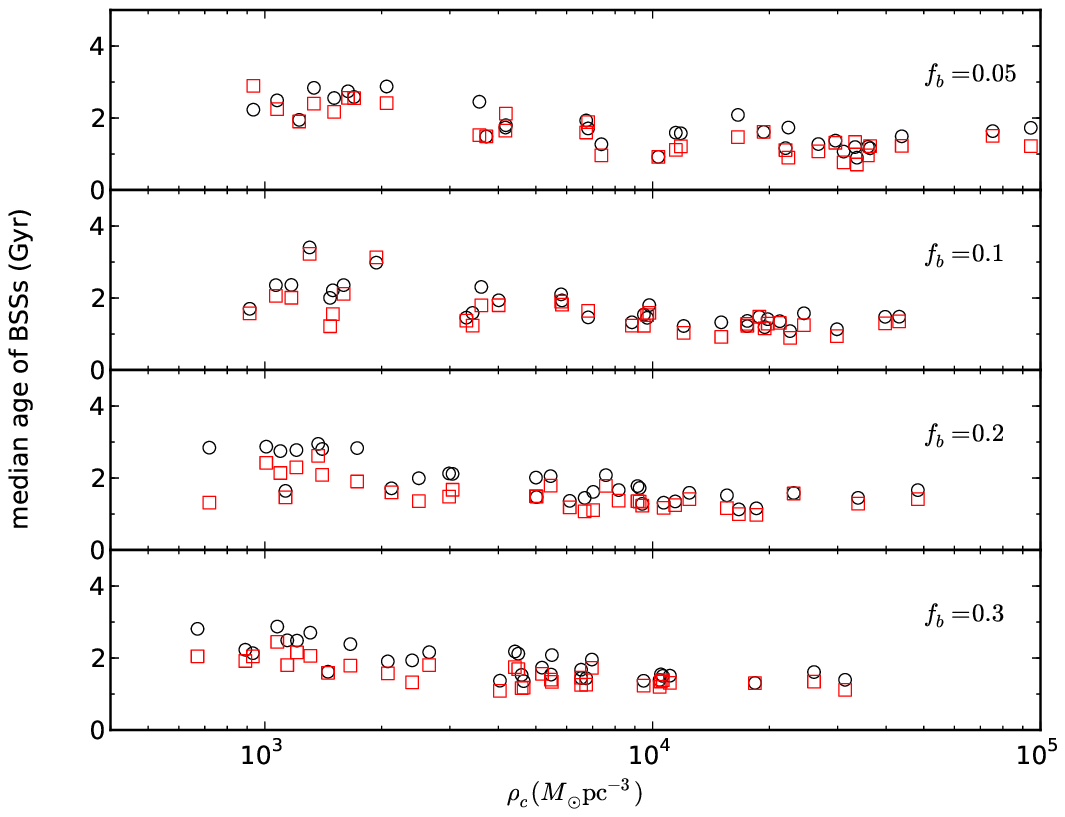}
\caption{The median values for $\tage$ of the BSSs in our models vs $\rho_c$.  Black circles and red squares denote 
all BSSs and BSSs in the core, respectively.  Top to bottom the panels denote models with initial $f_b = 0.05$, 
$0.1$, $0.2$, and $0.3$, respectively.  The median $\tage$ values for all initial $f_b$ show an anti-correlation with 
$\rho_c$.  The denser the core, the higher is the contribution from the collisional channels towards BSS production 
and the lower is the median $\tage$ of the BSSs.  
}
\label{plot:age_median_rhoc}
\end{center}
\end{figure*} 
The difference in the peaks of the age distributions between the high and low-density 
models is expected.  In a high-density model the dominant formation channel for BSSs is 
stellar collisions (\sing, \bs, and \bb; Figures\ \ref{plot:pie}, \ref{plot:rhoc_branching}).  
Contribution from \mtb\ increases as $\rho_c$ decreases (Figure\ \ref{plot:rhoc_branching}).
Stellar collisions typically form higher mass BSSs than 
those formed via \mtb.  Since a lower mass BSS has a longer residual lifetime 
than a higher mass counterpart, the typical $\tage$ becomes shorter as $\rho_c$ increases 
due to higher contribution from collisional BSS-formation channels.  
Indeed we find that the median for the $\tage$ values for the BSSs in our 
models negatively correlate with $\rho_c$ (correlation coefficient $r_{\rho_c, t_{\rm{age}}} = -0.44$; 
Figure\ \ref{plot:age_median_rhoc}).    
Due to the same reason the median value of $\tage$ for 
the core BSSs is typically slightly lower compared to the median value of $\tage$ 
for all BSSs in a cluster.  This effect is 
more prominent for higher $f_b$ models.  The number of BSSs outside $r_c$ 
is low for low $f_b$ models. Hence, the difference between the median $\tage$ values 
for BSSs in the whole cluster and in the core decreases with decreasing $f_b$. 
We find that the oldest BSSs in some clusters can be almost as old as the respective clusters.   

\section{Number of Core Blue-straggler Stars vs Binary Fraction and $\Gamma$}
\label{sec:gamma}
After the large-scale surveys using the Hubble Space Telescope \citep[e.g., the ACS survey;][]{2007AJ....133.1658S,2007AJ....134..376D,2008ApJ...673..241M} 
provided high resolution, homogeneously observed and analyzed data for a large number of GGCs, 
a number of studies have investigated various correlations between the number of BSSs in these clusters 
and their global properties 
\citep[e.g.,][]{2007ApJ...661..210L,2008A&A...481..701S,2009Natur.457..288K,2013MNRAS.428..897L}.  
In particular, the correlations or lack thereof between $\nbssc$ and 
two cluster properties namely the binary fraction in the core ($f_{b,c}$) and the collisional $\Gamma$ 
generated a lot of interest.  Both of these quantities are of great dynamical importance.  The presence 
of primordial binaries increases the BSS formation rates through all channels except \sing.  Instead, 
only BSS formation via \sing\ is directly related to the collisional $\Gamma$ given by 
\begin{equation}
\Gamma = \left( \frac{\rho_c}{\dens} \right)^2 \left( \frac{r_c}{\rm{pc}} \right)^3 \left( \frac{v_{c,\sigma}}{km s^{-1}} \right)^{-1}
\label{eq:gamma}
\end{equation}
\citep{2006ApJ...646L.143P}.  After it was found that the numbers 
of low-mass X-ray binaries and CVs in a cluster are strongly correlated with $\Gamma$, 
\citep[e.g.,][]{2006ApJ...646L.143P} a similar correlation was 
sought for $\nbssc$ since production of BSSs in the core was also expected to be 
affected significantly by dynamics. 
However, it was found that $\nbssc$ 
is rather insensitive to the global $\Gamma$ using data from HST/WFPC2 survey \citep{2002yCat..33910945P}.  
In contrast, it is found that $\nbssc$ shows clear correlation with the number of core binaries 
in the GGCs \citep[e.g.,][]{2013MNRAS.428..897L}.  
These two pieces of information were inferred as low importance of collisional channels in creation of 
BSSs in these clusters 
\citep[e.g.,][]{2009Natur.457..288K,2013MNRAS.428..897L}.  However, our results suggest that 
collisions dominate BSS production in the majority of GGCs.  We now investigate further this apparent 
discrepancy.    

\begin{figure*}
\begin{center}
\plotone{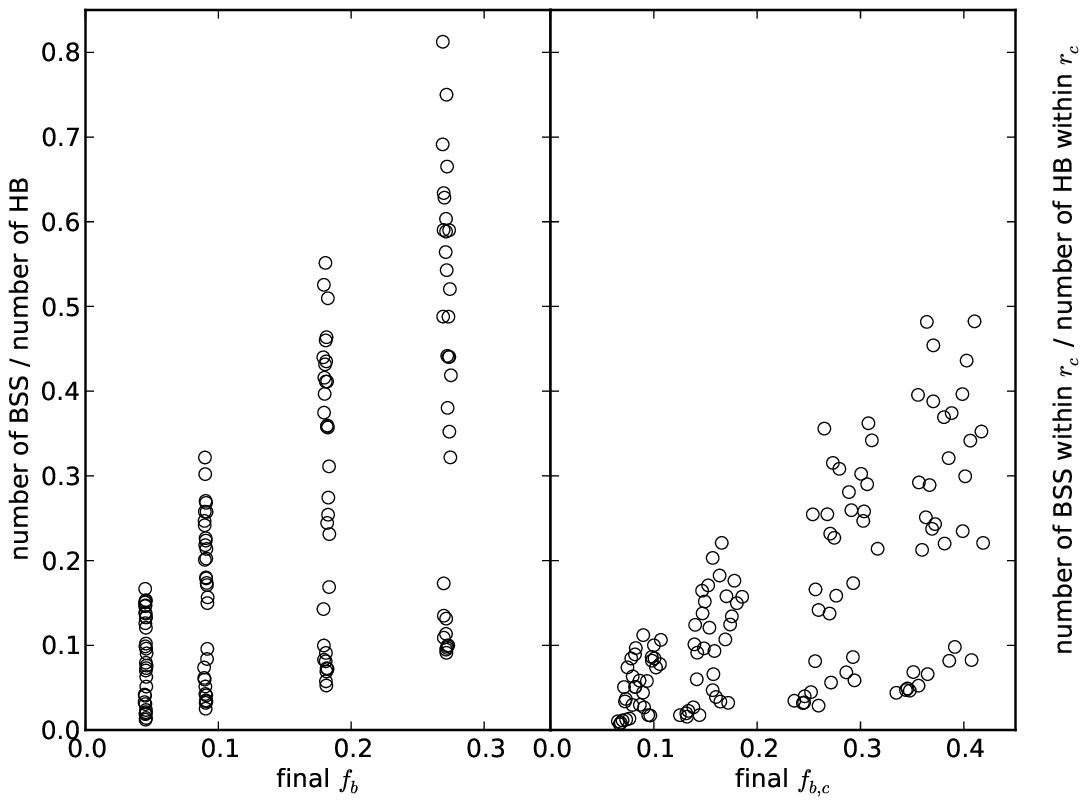}
\caption{Left Panel: The ratio of the number of BSSs ($\nbss$) to the number of horizontal-branch stars ($\nhb$) vs the final $f_b$.  
Right Panel: The ratio of the number of core BSSs ($\nbssc$) to the number of core horizontal-branch stars ($\nhbc$) 
vs the final core binary fraction ($f_{b,c}$).  The specific BSS frequencies $\nbss/\nhb$ and $\nbssc/\nhbc$ are 
correlated with $f_b$ and $f_{b,c}$, respectively.  
Although the dominant BSS production channel for these models is stellar collisions, most of these collisions 
are binary mediated.  Hence, a higher binary fraction leads to a higher number of collisions and thus a higher 
number of BSSs if all else is kept unchanged.  }
\label{plot:nbss_fb}
\end{center}
\end{figure*} 
Figure\ \ref{plot:nbss_fb} shows the specific number of BSSs ($\nbss/\nhb$) in our models as a function 
of the overall $f_b$.  It also shows $\nbssc/\nhbc$ as a function of the core binary 
fraction $f_{b,c}$ ($\nhb$ and $\nhbc$ are the numbers of horizontal branch stars in the cluster and in the core, respectively). 
Clearly, $\nbss/\nhb$ is correlated with $f_b$ and $\nbssc/\nhbc$ is 
correlated with $f_{b,c}$ (spearman correlation coefficient $r_{f_{b,c}, \nbssc} = 0.67$). 
For a more detailed comparison with observations see \citep{2013arXiv1303.2667S}. 
This is not at all surprising.  Results in Section\ \ref{sec:branching} clearly shows 
that throughout the studied range in $\rho_c$ the dominant BSS production channel is binary-mediated 
collisions (\bs\ and \bb).  Hence, as $f_b$ increases so does the number of BSSs.  Although, naturally the number of 
BSSs from the \mtb\ and \se\ channels also increases with increasing $f_b$, their relative importance is 
always low compared to \bs\ and \bb\ channels for all $\rho_c > 10^3\,\dens$.  Thus the observed correlation 
between $\nbssc/\nhbc$ with $f_{b,c}$ does not actually indicate BSS formation predominantly 
via \mtb, rather that most collisions are binary-mediated.         

\begin{figure*}
\begin{center}
\plotone{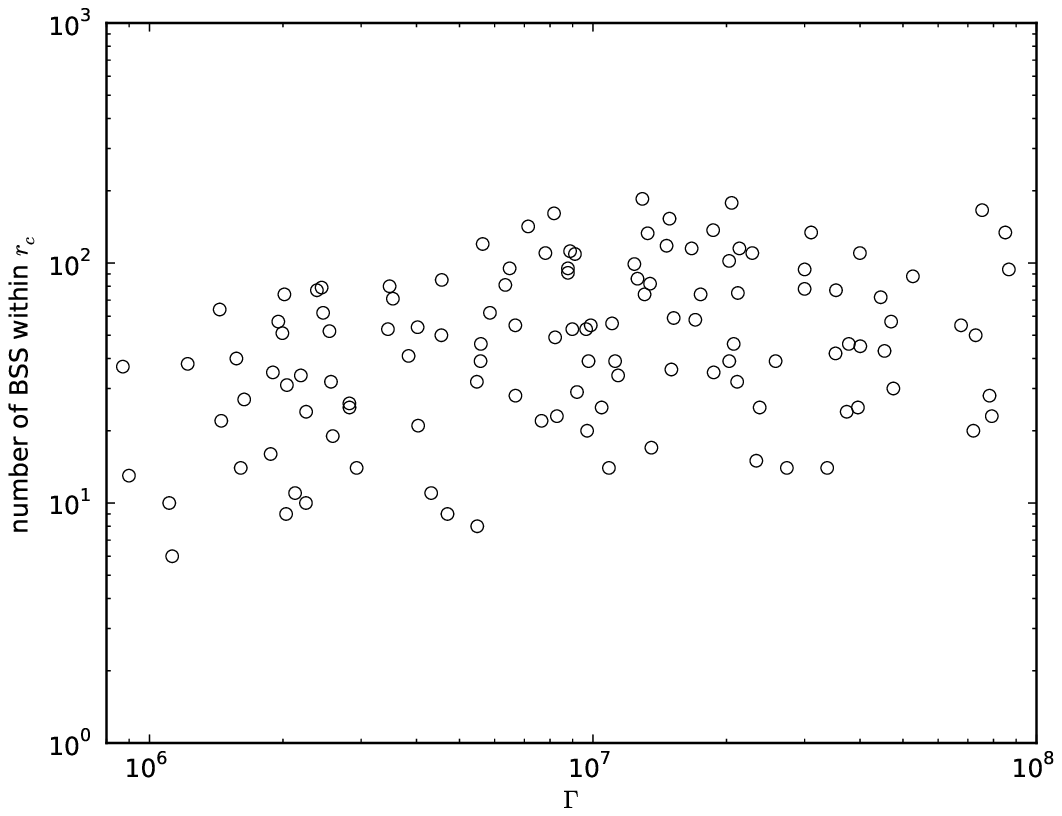}
\caption{$\nbssc$ vs $\Gamma$ \citep{2006ApJ...646L.143P} calculated 
using theoretical definitions in our models using Equation\ \ref{eq:gamma}.   
A weak positive correlation (correlation coefficient 0.12) between $\nbssc$ and $\Gamma$ is found.  }  
\label{plot:gamma_actual}
\end{center}
\end{figure*} 
Figure\ \ref{plot:gamma_actual} shows $\nbssc$ as a function of $\Gamma$ calculated using theoretical cluster properties 
in the code.  We find a weak positive correlation  
between $\nbssc$ and $\Gamma$ (spearman correlation coefficient $r_{\Gamma, \nbssc} = 0.12$). 

The positive correlations of $\nbssc$ with $f_{b,c}$ and $\Gamma$ are qualitatively consistent with the new ACS data 
\citep[][correlation coefficients are 0.4, and 0.5, respectively]{2011MNRAS.415.3771L}. \footnote{Note that the 
absolute values of the correlation coefficients from models and observed GGCs should not be compared and the comparison 
here is only qualitative. The unknown priors for the initial conditions 
used in these models may change these values. In addition, a more stringent comparison requires more observationally 
motivated determination of BSSs and cluster properties to find $\nbss$, $\nbssc$, $f_b$, $f_{b,c}$, and $\Gamma$. } 

Other exotic stellar populations, such as low-mass X ray binaries and cataclysmic variables are thought to be 
created when compact objects such as neutron stars and white dwarfs collide with giant stars, or acquire 
close binary companions via binary-mediated exchange interactions or tidal captures. Both populations show much tighter 
correlations with $\Gamma$ compared to $\nbssc$ \citep{2003ApJ...591L.131P,2006ApJ...646L.143P}. 
We expect this is because binaries with only a narrow range 
of properties (for example, one star must be a MS star, and the collision product must have mass higher than the 
$M_{\rm{TO}}$) create BSSs via collisions. 
In addition, the collision needs to occur at sufficiently late times such that the generated BSS is still in its residual H-burning stage when the cluster is observed.
BSSs 
produced via MTB, in particular for low-density clusters ($\rho_c < 10^3\,\dens$), also contributes to dilution of the 
correlation between $\nbssc$ and $\Gamma$. 
Mass segregation also reduces the number of MS stars in the core 
compared to relatively massive compact objects at late stages. Hence, production of those populations is more directly related 
to $\Gamma$ compared to production of BSSs.     

\section{Conclusions}
\label{sec:conclude}
We have identified the BSSs from a large set of (128) realistic cluster models 
using a simple and well-defined mass-based criterion. These models use observationally motivated initial 
conditions, include all relevant physics, and attain properties at $\sim 12\,\gyr$ including $N$, $f_b$, $\rho_c$, 
and $r_c$ typical of observed GGCs. In addition to cluster dynamical properties, all stellar properties are 
followed in tandem. For the first time these models allow us to investigate directly the full dynamical and 
stellar evolution history of each BSS in clusters spanning a large range of parameters. 
The BSSs identified in these models 
populate the expected region in a synthetic HRD (Figure\ \ref{plot:hrd}).  
In addition, the total numbers, radial distributions (Section \ref{sec:comparison}), and trends with various 
cluster parameters \citep[Section\ \ref{sec:gamma}; for more detailed comparisons see][]{2013arXiv1303.2667S} 
are all consistent with the latest observations of BSSs \citep{2011MNRAS.415.3771L}. 

We have identified the relative importance of the various relevant formation channels based on the detailed histories of all BSSs formed in our models.
We find that 
collisions play a major role in BSS formation for a large range of central densities typical of 
GGCs ($\rho_c \gtrsim 10^3\,\dens$; Figures\ \ref{plot:pie} -- \ref{plot:rhoc_branching_core}).  
BSSs in lower density clusters ($\rho_c \lesssim 10^3\,\dens$) can 
have a significant contribution from the \mtb\ channel (up to $\approx 60\%$ in some of our models).  This 
also indicates that for open cluster densities (typically $< 10^2\,\rm{\msun pc^{-3}}$) 
the dominant channel of BSS formation should be MTB. Binaries are very important for significant 
production of BSSs in all clusters since most collisions happen via binary-mediated interactions 
(Figures\ \ref{plot:rhoc_branching_coll}, \ref{plot:rhoc_branching_coll_core}). Even among BSSs created via 
binary stellar evolution channels (\mtb\ and \se), for $\rho_c\gtrsim10^4\,\dens$ more than $50\%$ may come from 
binaries that have been perturbed via dynamical interactions, thereby changing their evolutionary pathways before 
BSS production.     
This basic picture could only be changed if, in reality, the degree of mixing from the \mtb\ or \se\ channels 
were very different than what is expected from collisions. As an extreme example, if collisions led to 
no significant mixing, with no fresh Hydrogen supplied to the stellar core, collisions would produce BSSs 
with much shorter lifetimes. This might increase the relative contribution of \mtb\ in producing the 
observed BSSs if 
\mtb\ leads to more mixing and a higher level of rejuvenation. However, as long as 
collisional channels and \mtb\ produce similar levels of rejuvenation, our results remain 
at least qualitatively valid.  

The importance of binary-mediated collisions in the production of BSSs 
is of great significance.  On the one hand this explains the observed correlation between $\nbssc$ and 
$f_{b,c}$, in particular, in the higher-density ($\rho_c>10^3\,\dens$) GGCs. 
On the other hand, the importance of collisions suggests that there should be some correlation 
between $\nbssc$ and the global $\Gamma$. Indeed, 
we find that $\nbssc$ and $\Gamma$ do show some correlation (although weak) for 
both our models and the GGCs based on the new HST/ACS data with the best available photometry. 
Early studies \citep[e.g.,][]{2004MNRAS.349..129D,2007ApJ...661..210L} 
that claimed no correlation (or anti-correlation) between $\nbssc$ and $\Gamma$ 
used HST/WFPC2 data with lower photometric precision. 

Throughout this study we have adopted a simple full-mixing prescription for rejuvenation. 
Based on this assumption we 
estimate upper limits for typical $t_{\rm{age}}$ for these models. 
In our models, the oldest BSSs can be almost as old as the host clusters.  
The typical median ages of the BSSs are, however, between $\approx 1$ and $3\,\rm{Gyr}$ 
(depending on $\rho_c$; Figure\ \ref{plot:age_median_rhoc} and Table\ \ref{tab:runlist}). 
This is consistent with previous estimates for observed BSSs in 47-Tuc \citep{2000ApJ...535..298S}.  
Thus, present-day BSSs are possibly affected by the dynamical history of their host clusters 
during the past few billion years. Hence, rate-based studies should take into account the changing global 
cluster properties during the last $\sim t_{\rm{age}}$. Future studies should improve this estimate using more 
realistic rejuvenation prescriptions via, for example, live single and binary stellar evolution codes that can 
work in parallel with $N$-body codes. However, although, $t_{\rm{age}}$ is likely to decrease \citep[e.g.,][]{2008A&A...488.1017G}, 
we expect the branching ratios for the various formation channels to remain similar.   

\section{Acknowledgements} We thank C.~Knigge, N.~Leigh, and an anonymous referee for valuable suggestions. 
Support for this study was provided by NASA through a grant from the Space Telescope 
Science Institute, which is operated by the Association of Universities for Research in Astronomy, Inc., under 
NASA contract NAS5-26555; the grant identifying number is HST-AR-12829.01-A.  SC also acknowledges support 
from the Department of Astronomy at the University of Florida. FAR acknowledges support from NASA ATP Grant 
NNX09AO36G at Northwestern University and from NSF Grant PHY-1066293 at the Aspen Center for Physics. 
AS is supported by the Natural Sciences and Engineering Research Council of Canada.  
All authors would like to thank the Kavli Institute for Theoretical Physics, where this project began; this work 
was supported in part at KITP by the National Science Foundation under Grant PHY11-25915.

\clearpage
\LongTables
\begin{deluxetable}{l|cccccccc|cccccccccc|ccc}
\tabletypesize{\footnotesize}
\tablecolumns{21}
\tablewidth{0pt}
\tablecaption{Model properties and dynamical properties of BSSs. \label{tab:runlist} 
}
\tablehead{
\multicolumn{1}{c}{Name} & 
\multicolumn{1}{c}{$t_{\rm{cl}}$} & 
\multicolumn{1}{c}{$M$} & 
\multicolumn{1}{c}{$f_b$} & 
\multicolumn{1}{c}{$f_{b,c}$} & 
\multicolumn{1}{c}{$\rho_c$} & 
\multicolumn{1}{c}{$\rho_{c,\rm{obs}}$} &
\multicolumn{1}{c}{$r_c$} & 
\multicolumn{1}{c}{$r_{c,\rm{obs}}$} & 
\multicolumn{5}{c}{$N_{\rm{BSS}}$} & 
\multicolumn{5}{c}{$N_{\rm{BSS},c}$} & 
\multicolumn{2}{c}{$t_{\rm{age}}$} \\ \cline{10-14} \cline{15-19} \cline{20-21}
\multicolumn{1}{c}{} &
\multicolumn{1}{c}{} &
\multicolumn{1}{c}{} &
\multicolumn{1}{c}{} &
\multicolumn{1}{c}{} &
\multicolumn{2}{c}{} &
\multicolumn{2}{c}{} &
\multicolumn{3}{c}{Coll} &
\multicolumn{1}{c}{MTB} &
\multicolumn{1}{c}{SE} &
\multicolumn{3}{c}{Coll} &
\multicolumn{1}{c}{MTB} &
\multicolumn{1}{c}{SE} &
\multicolumn{1}{c}{Med} &
\multicolumn{1}{c}{Max} \\ \cline{10-12} \cline{15-17}
\multicolumn{1}{c}{} &
\multicolumn{1}{c}{} &
\multicolumn{1}{c}{} &
\multicolumn{1}{c}{} &
\multicolumn{1}{c}{} &
\multicolumn{1}{c}{} &
\multicolumn{1}{c}{} &
\multicolumn{1}{c}{} &
\multicolumn{1}{c}{} &
\multicolumn{1}{c}{{\tt s-s}} &
\multicolumn{1}{c}{\bs} &
\multicolumn{1}{c}{\bb} &
\multicolumn{1}{c}{} &
\multicolumn{1}{c}{} &
\multicolumn{1}{c}{{\tt s-s}} &
\multicolumn{1}{c}{\bs} &
\multicolumn{1}{c}{\bb} &
\multicolumn{1}{c}{} &
\multicolumn{1}{c}{} &
\multicolumn{2}{c}{} 
}
\startdata
%
{\tt run0} & 12 & 14 & 0.05 & 0.09 & 1.03 & 0.29 & 0.95 & 1.43 & 2 & 11 & 3 & 1 & 0 & 1 & 10 & 3 & 0 & 0 & 0.9 & 5.2 \\
{\tt run2} & 12 & 21 & 0.05 & 0.09 & 0.74 & 0.09 & 1.23 & 2.97 & 4 & 24 & 4 & 2 & 1 & 3 & 15 & 2 & 0 & 0 & 1.3 & 5.9 \\
{\tt run3} & 12 & 28 & 0.05 & 0.08 & 0.68 & 37.00 & 1.39 & 0.08 & 6 & 19 & 10 & 3 & 1 & 2 & 15 & 5 & 3 & 0 & 1.7 & 5.4 \\
{\tt run4} & 12 & 36 & 0.05 & 0.07 & 0.67 & 0.89 & 1.50 & 1.41 & 11 & 30 & 7 & 10 & 0 & 7 & 17 & 5 & 5 & 0 & 1.9 & 7.3 \\
{\tt run5} & 12 & 14 & 0.09 & 0.17 & 0.88 & 0.20 & 1.02 & 1.93 & 2 & 20 & 27 & 1 & 0 & 1 & 17 & 20 & 1 & 0 & 1.3 & 7.3 \\
{\tt run6} & 12 & 22 & 0.09 & 0.16 & 0.68 & 0.97 & 1.27 & 1.00 & 4 & 24 & 19 & 8 & 1 & 1 & 14 & 10 & 4 & 0 & 1.5 & 5.4 \\
{\tt run7} & 12 & 29 & 0.09 & 0.15 & 0.58 & 1.02 & 1.46 & 1.14 & 5 & 45 & 24 & 10 & 0 & 1 & 32 & 16 & 4 & 0 & 1.9 & 6.8 \\
{\tt run8} & 12 & 36 & 0.09 & 0.14 & 0.58 & 0.22 & 1.58 & 2.66 & 10 & 38 & 25 & 15 & 1 & 5 & 25 & 18 & 6 & 1 & 2.1 & 7.3 \\
{\tt run9} & 12 & 15 & 0.18 & 0.29 & 0.50 & 0.85 & 1.23 & 0.82 & 1 & 17 & 30 & 5 & 2 & 1 & 13 & 20 & 5 & 0 & 1.5 & 10.0 \\
{\tt run10} & 12 & 23 & 0.18 & 0.29 & 0.61 & 35.43 & 1.34 & 0.09 & 1 & 49 & 73 & 15 & 3 & 0 & 37 & 46 & 7 & 1 & 1.4 & 5.5 \\
{\tt run11} & 12 & 30 & 0.18 & 0.27 & 0.55 & 0.45 & 1.52 & 1.73 & 4 & 50 & 78 & 28 & 1 & 3 & 31 & 51 & 9 & 1 & 2.1 & 8.6 \\
{\tt run12} & 12 & 38 & 0.18 & 0.25 & 0.50 & 0.19 & 1.68 & 2.70 & 3 & 75 & 60 & 44 & 1 & 2 & 52 & 40 & 18 & 0 & 2.0 & 8.1 \\
{\tt run13} & 12 & 16 & 0.27 & 0.40 & 0.40 & 0.20 & 1.32 & 1.85 & 0 & 26 & 45 & 8 & 2 & 0 & 13 & 33 & 3 & 1 & 1.4 & 7.3 \\
{\tt run14} & 12 & 24 & 0.27 & 0.39 & 0.46 & 0.52 & 1.47 & 1.39 & 3 & 44 & 99 & 17 & 4 & 0 & 25 & 61 & 8 & 1 & 1.5 & 9.1 \\
{\tt run15} & 12 & 32 & 0.27 & 0.37 & 0.44 & 0.46 & 1.64 & 1.59 & 5 & 64 & 115 & 38 & 8 & 2 & 41 & 81 & 16 & 2 & 2.2 & 10.0 \\
{\tt run16} & 12 & 40 & 0.27 & 0.36 & 0.45 & 0.18 & 1.76 & 2.98 & 2 & 97 & 106 & 49 & 4 & 1 & 63 & 76 & 19 & 2 & 2.1 & 12.0 \\
{\tt run17} & 12 & 14 & 0.05 & 0.11 & 3.36 & 54.35 & 0.67 & 0.09 & 6 & 22 & 8 & 0 & 0 & 4 & 15 & 6 & 0 & 0 & 0.9 & 5.6 \\
{\tt run18} & 12 & 21 & 0.05 & 0.09 & 1.66 & 50.28 & 0.95 & 0.07 & 2 & 14 & 9 & 1 & 1 & 0 & 5 & 9 & 0 & 1 & 2.1 & 7.6 \\
{\tt run19} & 12 & 28 & 0.04 & 0.08 & 1.18 & 0.33 & 1.17 & 2.00 & 10 & 38 & 7 & 1 & 3 & 4 & 24 & 5 & 0 & 2 & 1.6 & 5.9 \\
{\tt run20} & 12 & 36 & 0.04 & 0.08 & 1.15 & 0.11 & 1.27 & 3.43 & 12 & 36 & 12 & 2 & 2 & 4 & 25 & 8 & 1 & 1 & 1.6 & 7.2 \\
{\tt run21} & 12 & 14 & 0.09 & 0.19 & 2.13 & 0.49 & 0.78 & 1.05 & 1 & 24 & 26 & 1 & 1 & 0 & 19 & 20 & 0 & 0 & 1.4 & 5.5 \\
{\tt run22} & 12 & 22 & 0.09 & 0.16 & 1.20 & 0.22 & 1.06 & 2.18 & 5 & 49 & 29 & 1 & 2 & 1 & 33 & 22 & 1 & 1 & 1.2 & 7.0 \\
{\tt run23} & 12 & 29 & 0.09 & 0.15 & 0.95 & 0.36 & 1.26 & 1.83 & 10 & 51 & 24 & 9 & 2 & 5 & 30 & 17 & 6 & 1 & 1.5 & 10.9 \\
{\tt run24} & 12 & 36 & 0.09 & 0.15 & 0.98 & 0.94 & 1.34 & 1.36 & 10 & 61 & 35 & 8 & 2 & 6 & 41 & 23 & 2 & 2 & 1.8 & 9.2 \\
{\tt run25} & 12 & 15 & 0.18 & 0.30 & 0.94 & 0.69 & 1.01 & 0.99 & 1 & 25 & 52 & 0 & 0 & 0 & 16 & 40 & 0 & 0 & 1.3 & 7.4 \\
{\tt run26} & 12 & 23 & 0.18 & 0.29 & 0.91 & 2.27 & 1.18 & 0.73 & 2 & 53 & 61 & 13 & 1 & 0 & 36 & 41 & 4 & 1 & 1.8 & 11.5 \\
{\tt run27} & 12 & 30 & 0.18 & 0.27 & 0.76 & 120.61 & 1.37 & 0.10 & 3 & 48 & 59 & 27 & 2 & 2 & 35 & 35 & 13 & 1 & 2.1 & 9.5 \\
{\tt run28} & 12 & 38 & 0.18 & 0.26 & 0.82 & 0.56 & 1.44 & 1.71 & 2 & 83 & 110 & 28 & 3 & 0 & 57 & 83 & 12 & 1 & 1.7 & 6.4 \\
{\tt run29} & 12 & 16 & 0.27 & 0.40 & 0.52 & 1.98 & 1.22 & 0.52 & 0 & 29 & 62 & 8 & 2 & 0 & 15 & 41 & 5 & 1 & 1.7 & 6.8 \\
{\tt run30} & 12 & 24 & 0.27 & 0.39 & 0.55 & 0.33 & 1.39 & 1.67 & 0 & 58 & 98 & 11 & 6 & 0 & 35 & 71 & 4 & 0 & 1.5 & 7.5 \\
{\tt run31} & 12 & 32 & 0.27 & 0.37 & 0.55 & 0.81 & 1.53 & 1.16 & 1 & 54 & 96 & 29 & 4 & 1 & 35 & 62 & 9 & 2 & 2.1 & 8.0 \\
{\tt run32} & 12 & 40 & 0.27 & 0.36 & 0.70 & 7.71 & 1.53 & 0.37 & 1 & 112 & 156 & 33 & 10 & 0 & 66 & 107 & 11 & 1 & 2.0 & 11.1 \\
{\tt run33*} & 11 & 14 & 0.05 & 0.10 & 3.33 & 379.65 & 0.66 & 0.06 & 3 & 19 & 8 & 2 & 1 & 2 & 15 & 6 & 0 & 1 & 1.2 & 5.2 \\
{\tt run34} & 11 & 21 & 0.05 & 0.10 & 3.11 & 4.46 & 0.79 & 0.61 & 6 & 24 & 10 & 2 & 0 & 3 & 16 & 9 & 2 & 0 & 1.1 & 3.9 \\
{\tt run35} & 11 & 29 & 0.04 & 0.09 & 2.67 & 3.25 & 0.91 & 0.67 & 9 & 43 & 10 & 1 & 1 & 3 & 29 & 10 & 0 & 1 & 1.3 & 5.8 \\
{\tt run36} & 11 & 36 & 0.04 & 0.08 & 2.20 & 1.50 & 1.03 & 1.17 & 12 & 39 & 13 & 1 & 3 & 6 & 28 & 8 & 0 & 3 & 1.2 & 5.2 \\
{\tt run37} & 11 & 14 & 0.09 & 0.17 & 1.75 & 59.93 & 0.82 & 0.07 & 2 & 21 & 21 & 2 & 0 & 1 & 15 & 15 & 1 & 0 & 1.4 & 6.5 \\
{\tt run38} & 11 & 22 & 0.09 & 0.18 & 2.26 & 0.49 & 0.89 & 1.56 & 2 & 47 & 19 & 1 & 1 & 1 & 30 & 11 & 0 & 0 & 1.1 & 5.7 \\
{\tt run39} & 11 & 29 & 0.09 & 0.16 & 1.75 & 2.44 & 1.05 & 0.76 & 6 & 62 & 37 & 5 & 6 & 1 & 44 & 27 & 1 & 5 & 1.2 & 6.8 \\
{\tt run40} & 11 & 37 & 0.09 & 0.15 & 1.88 & 91.50 & 1.09 & 0.08 & 7 & 60 & 27 & 12 & 3 & 3 & 42 & 23 & 7 & 2 & 1.5 & 9.1 \\
{\tt run41} & 11 & 15 & 0.18 & 0.32 & 1.67 & 0.55 & 0.85 & 1.44 & 0 & 28 & 47 & 1 & 1 & 0 & 18 & 28 & 0 & 0 & 1.1 & 5.6 \\
{\tt run42} & 11 & 23 & 0.18 & 0.31 & 1.85 & 0.83 & 0.96 & 1.07 & 5 & 61 & 74 & 8 & 1 & 1 & 38 & 50 & 5 & 0 & 1.2 & 8.8 \\
{\tt run43} & 11 & 30 & 0.18 & 0.28 & 1.24 & 1.15 & 1.17 & 1.08 & 3 & 52 & 81 & 10 & 2 & 2 & 40 & 67 & 4 & 2 & 1.6 & 6.0 \\
{\tt run44} & 12 & 38 & 0.18 & 0.27 & 1.55 & 1.23 & 1.19 & 1.23 & 6 & 85 & 80 & 12 & 4 & 5 & 62 & 60 & 4 & 3 & 1.5 & 7.1 \\
{\tt run45} & 12 & 15 & 0.27 & 0.42 & 1.11 & 1.15 & 0.95 & 0.73 & 2 & 30 & 76 & 3 & 3 & 0 & 20 & 53 & 1 & 0 & 1.5 & 7.2 \\
{\tt run46} & 11 & 24 & 0.27 & 0.40 & 1.04 & 4.31 & 1.15 & 0.45 & 2 & 53 & 103 & 10 & 7 & 1 & 33 & 74 & 4 & 3 & 1.4 & 6.8 \\
{\tt run47} & 12 & 32 & 0.27 & 0.38 & 1.06 & 0.97 & 1.25 & 1.09 & 4 & 73 & 127 & 11 & 4 & 1 & 46 & 85 & 4 & 1 & 1.5 & 10.0 \\
{\tt run48} & 12 & 40 & 0.27 & 0.37 & 1.05 & 0.51 & 1.35 & 1.69 & 3 & 77 & 161 & 26 & 4 & 2 & 53 & 115 & 6 & 2 & 1.5 & 9.6 \\
{\tt run49} & 12 & 14 & 0.05 & 0.07 & 0.09 & 0.19 & 2.11 & 1.72 & 0 & 8 & 1 & 4 & 0 & 0 & 4 & 1 & 1 & 0 & 2.2 & 5.2 \\
{\tt run50} & 12 & 21 & 0.05 & 0.07 & 0.11 & 0.03 & 2.32 & 4.19 & 1 & 9 & 3 & 7 & 1 & 1 & 5 & 2 & 5 & 1 & 2.5 & 7.5 \\
{\tt run51} & 12 & 29 & 0.05 & 0.07 & 0.12 & 0.06 & 2.45 & 3.95 & 5 & 10 & 0 & 8 & 0 & 1 & 6 & 0 & 4 & 0 & 1.9 & 7.4 \\
{\tt run52} & 12 & 36 & 0.05 & 0.07 & 0.13 & 0.20 & 2.56 & 2.49 & 3 & 15 & 1 & 20 & 0 & 3 & 9 & 1 & 6 & 0 & 2.8 & 7.3 \\
{\tt run53} & 12 & 14 & 0.09 & 0.14 & 0.09 & 0.04 & 2.14 & 3.29 & 1 & 5 & 6 & 4 & 0 & 0 & 4 & 5 & 1 & 0 & 1.7 & 4.9 \\
{\tt run54} & 12 & 22 & 0.09 & 0.14 & 0.11 & 0.10 & 2.35 & 2.54 & 1 & 14 & 12 & 13 & 0 & 0 & 8 & 10 & 9 & 0 & 2.4 & 6.6 \\
{\tt run55} & 12 & 29 & 0.09 & 0.13 & 0.12 & 0.03 & 2.49 & 4.70 & 5 & 22 & 12 & 14 & 0 & 2 & 15 & 7 & 7 & 0 & 2.4 & 7.5 \\
{\tt run56} & 12 & 37 & 0.09 & 0.13 & 0.13 & 2.82 & 2.60 & 0.31 & 3 & 21 & 6 & 33 & 1 & 2 & 12 & 4 & 14 & 0 & 3.4 & 8.7 \\
{\tt run57} & 12 & 15 & 0.18 & 0.26 & 0.07 & 2.43 & 2.34 & 0.17 & 0 & 7 & 9 & 11 & 0 & 0 & 5 & 5 & 3 & 0 & 2.8 & 6.1 \\
{\tt run58} & 12 & 23 & 0.18 & 0.26 & 0.10 & 0.16 & 2.41 & 2.13 & 0 & 32 & 19 & 23 & 1 & 0 & 18 & 14 & 7 & 1 & 2.9 & 7.4 \\
{\tt run59} & 12 & 31 & 0.18 & 0.25 & 0.11 & 0.20 & 2.59 & 2.02 & 0 & 29 & 32 & 51 & 1 & 0 & 19 & 19 & 13 & 0 & 2.7 & 10.4 \\
{\tt run60} & 12 & 38 & 0.18 & 0.24 & 0.12 & 0.04 & 2.70 & 5.07 & 5 & 30 & 21 & 67 & 1 & 3 & 21 & 14 & 24 & 0 & 2.8 & 10.4 \\
{\tt run61} & 12 & 16 & 0.27 & 0.36 & 0.07 & 5.88 & 2.43 & 0.17 & 1 & 9 & 20 & 26 & 0 & 1 & 7 & 14 & 15 & 0 & 2.8 & 5.6 \\
{\tt run62} & 12 & 24 & 0.27 & 0.36 & 0.09 & 3.78 & 2.55 & 0.17 & 0 & 34 & 41 & 36 & 3 & 0 & 22 & 24 & 18 & 0 & 2.2 & 7.3 \\
{\tt run63} & 12 & 32 & 0.27 & 0.35 & 0.11 & 0.12 & 2.64 & 2.99 & 1 & 42 & 41 & 73 & 2 & 0 & 25 & 24 & 24 & 1 & 2.9 & 11.8 \\
{\tt run64} & 12 & 40 & 0.27 & 0.33 & 0.11 & 0.06 & 2.78 & 4.07 & 1 & 41 & 49 & 77 & 2 & 0 & 24 & 34 & 19 & 0 & 2.5 & 10.6 \\
{\tt run65} & 12 & 14 & 0.05 & 0.08 & 0.16 & 0.09 & 1.77 & 2.17 & 3 & 6 & 4 & 3 & 0 & 2 & 2 & 4 & 1 & 0 & 2.7 & 5.2 \\
{\tt run66} & 12 & 21 & 0.05 & 0.07 & 0.15 & 2.84 & 2.07 & 0.25 & 1 & 7 & 0 & 6 & 0 & 1 & 7 & 0 & 2 & 0 & 2.6 & 6.3 \\
{\tt run67} & 12 & 29 & 0.05 & 0.07 & 0.17 & 0.07 & 2.19 & 3.45 & 3 & 8 & 1 & 8 & 0 & 1 & 6 & 1 & 6 & 0 & 2.6 & 8.5 \\
{\tt run68} & 12 & 36 & 0.05 & 0.07 & 0.21 & 0.06 & 2.22 & 4.26 & 6 & 20 & 5 & 13 & 0 & 1 & 13 & 3 & 4 & 0 & 2.9 & 7.0 \\
{\tt run69} & 12 & 14 & 0.09 & 0.15 & 0.15 & 0.13 & 1.83 & 2.21 & 1 & 10 & 8 & 6 & 1 & 0 & 6 & 5 & 4 & 1 & 2.2 & 6.6 \\
{\tt run70} & 12 & 22 & 0.09 & 0.14 & 0.15 & 5.07 & 2.10 & 0.14 & 3 & 17 & 12 & 13 & 0 & 2 & 7 & 9 & 6 & 0 & 2.0 & 6.9 \\
{\tt run71} & 12 & 29 & 0.09 & 0.13 & 0.16 & 5.08 & 2.25 & 0.12 & 1 & 15 & 7 & 17 & 0 & 1 & 12 & 4 & 8 & 0 & 2.4 & 10.5 \\
{\tt run72} & 12 & 37 & 0.09 & 0.13 & 0.19 & 0.08 & 2.28 & 3.55 & 4 & 33 & 10 & 24 & 3 & 1 & 20 & 6 & 13 & 1 & 3.0 & 11.3 \\
{\tt run73} & 12 & 15 & 0.18 & 0.27 & 0.11 & 0.11 & 2.03 & 2.06 & 0 & 11 & 15 & 11 & 0 & 0 & 5 & 11 & 6 & 0 & 1.6 & 8.2 \\
{\tt run74} & 12 & 23 & 0.18 & 0.26 & 0.14 & 0.06 & 2.19 & 3.42 & 3 & 16 & 22 & 19 & 1 & 0 & 11 & 16 & 7 & 0 & 2.9 & 7.2 \\
{\tt run75} & 12 & 30 & 0.18 & 0.24 & 0.14 & 0.07 & 2.38 & 3.49 & 1 & 29 & 21 & 42 & 1 & 1 & 19 & 17 & 14 & 1 & 2.8 & 6.9 \\
{\tt run76} & 12 & 38 & 0.18 & 0.25 & 0.17 & 0.14 & 2.40 & 2.97 & 3 & 36 & 37 & 52 & 2 & 3 & 21 & 28 & 18 & 1 & 2.8 & 10.0 \\
{\tt run77} & 12 & 16 & 0.28 & 0.37 & 0.09 & 0.05 & 2.18 & 3.01 & 1 & 23 & 24 & 19 & 0 & 0 & 14 & 15 & 9 & 0 & 2.1 & 8.2 \\
{\tt run78} & 12 & 24 & 0.27 & 0.36 & 0.12 & 7.73 & 2.30 & 0.13 & 0 & 21 & 37 & 41 & 1 & 0 & 12 & 24 & 20 & 1 & 2.5 & 11.7 \\
{\tt run79} & 12 & 32 & 0.27 & 0.35 & 0.13 & 0.13 & 2.46 & 2.68 & 2 & 27 & 46 & 73 & 5 & 2 & 20 & 31 & 26 & 0 & 2.7 & 9.5 \\
{\tt run80} & 12 & 40 & 0.27 & 0.34 & 0.17 & 0.13 & 2.46 & 2.98 & 2 & 48 & 55 & 59 & 4 & 0 & 23 & 37 & 19 & 1 & 2.4 & 7.4 \\
{\tt run81} & 12 & 14 & 0.05 & 0.08 & 0.37 & 0.23 & 1.35 & 1.54 & 1 & 10 & 5 & 1 & 0 & 0 & 7 & 2 & 0 & 0 & 1.5 & 5.2 \\
{\tt run82} & 12 & 21 & 0.05 & 0.08 & 0.36 & 0.32 & 1.57 & 1.45 & 1 & 10 & 2 & 4 & 0 & 1 & 5 & 2 & 0 & 0 & 2.5 & 7.0 \\
{\tt run83} & 12 & 28 & 0.05 & 0.08 & 0.42 & 0.23 & 1.65 & 2.18 & 8 & 21 & 2 & 1 & 0 & 4 & 16 & 1 & 1 & 0 & 1.8 & 5.2 \\
{\tt run84} & 12 & 36 & 0.05 & 0.07 & 0.42 & 0.13 & 1.76 & 3.01 & 4 & 20 & 3 & 7 & 1 & 1 & 16 & 3 & 3 & 0 & 1.7 & 6.3 \\
{\tt run85} & 12 & 14 & 0.09 & 0.16 & 0.33 & 0.61 & 1.42 & 0.81 & 0 & 8 & 6 & 0 & 0 & 0 & 7 & 4 & 0 & 0 & 1.5 & 6.1 \\
{\tt run86} & 12 & 22 & 0.09 & 0.15 & 0.34 & 0.20 & 1.61 & 1.99 & 1 & 21 & 18 & 6 & 0 & 1 & 13 & 15 & 3 & 0 & 1.6 & 10.5 \\
{\tt run87} & 12 & 29 & 0.09 & 0.14 & 0.36 & 0.14 & 1.74 & 2.76 & 4 & 22 & 13 & 11 & 1 & 1 & 17 & 6 & 4 & 0 & 2.3 & 10.5 \\
{\tt run88} & 12 & 36 & 0.09 & 0.14 & 0.40 & 0.24 & 1.81 & 2.25 & 10 & 31 & 22 & 12 & 3 & 6 & 23 & 17 & 3 & 0 & 1.9 & 7.0 \\
{\tt run89} & 12 & 15 & 0.18 & 0.28 & 0.21 & 0.16 & 1.66 & 1.96 & 0 & 27 & 13 & 5 & 0 & 0 & 19 & 7 & 0 & 0 & 1.7 & 5.2 \\
{\tt run90} & 12 & 23 & 0.18 & 0.27 & 0.25 & 0.19 & 1.80 & 2.00 & 2 & 31 & 34 & 16 & 2 & 0 & 23 & 23 & 8 & 0 & 2.0 & 7.5 \\
{\tt run91} & 12 & 30 & 0.18 & 0.26 & 0.30 & 0.12 & 1.86 & 2.78 & 1 & 25 & 31 & 26 & 1 & 0 & 18 & 19 & 9 & 0 & 2.1 & 6.7 \\
{\tt run92} & 12 & 38 & 0.18 & 0.25 & 0.30 & 0.33 & 1.99 & 1.90 & 5 & 39 & 60 & 38 & 5 & 1 & 24 & 40 & 15 & 1 & 2.1 & 7.8 \\
{\tt run93} & 12 & 16 & 0.27 & 0.38 & 0.15 & 0.22 & 1.87 & 1.36 & 0 & 15 & 32 & 9 & 0 & 0 & 10 & 23 & 2 & 0 & 1.6 & 6.9 \\
{\tt run94} & 12 & 24 & 0.27 & 0.37 & 0.21 & 1.11 & 1.93 & 0.76 & 1 & 26 & 49 & 19 & 1 & 0 & 15 & 29 & 8 & 1 & 1.9 & 7.2 \\
{\tt run95} & 12 & 32 & 0.27 & 0.36 & 0.24 & 0.17 & 2.02 & 2.26 & 2 & 36 & 64 & 43 & 3 & 1 & 20 & 49 & 13 & 2 & 1.9 & 8.1 \\
{\tt run96} & 12 & 40 & 0.27 & 0.35 & 0.27 & 27.80 & 2.11 & 0.11 & 2 & 57 & 82 & 53 & 5 & 1 & 31 & 62 & 22 & 4 & 2.2 & 7.5 \\
{\tt run97*} & 12 & 14 & 0.05 & 0.11 & 2.24 & 0.27 & 0.79 & 1.36 & 1 & 12 & 11 & 0 & 0 & 0 & 7 & 7 & 0 & 0 & 1.7 & 6.8 \\
{\tt run98} & 12 & 21 & 0.05 & 0.10 & 3.59 & 11.10 & 0.79 & 0.24 & 3 & 18 & 15 & 0 & 0 & 0 & 10 & 10 & 0 & 0 & 1.2 & 7.5 \\
{\tt run99} & 12 & 28 & 0.05 & 0.10 & 3.64 & 65.46 & 0.85 & 0.07 & 4 & 24 & 19 & 1 & 2 & 2 & 16 & 10 & 0 & 0 & 1.2 & 4.3 \\
{\tt run100} & 12 & 36 & 0.04 & 0.09 & 2.96 & 1.33 & 0.96 & 0.95 & 7 & 53 & 13 & 2 & 0 & 6 & 40 & 9 & 0 & 0 & 1.4 & 6.2 \\
{\tt run101*} & 12 & 14 & 0.09 & 0.18 & 1.50 & 21.45 & 0.91 & 0.09 & 0 & 20 & 18 & 4 & 1 & 0 & 11 & 13 & 0 & 1 & 1.3 & 5.2 \\
{\tt run102*} & 12 & 22 & 0.09 & 0.18 & 1.98 & 1.15 & 0.95 & 0.81 & 5 & 31 & 30 & 3 & 1 & 3 & 20 & 21 & 1 & 1 & 1.4 & 6.0 \\
{\tt run103} & 12 & 29 & 0.09 & 0.17 & 2.99 & 1.96 & 0.94 & 0.80 & 2 & 41 & 34 & 1 & 2 & 1 & 27 & 22 & 0 & 0 & 1.1 & 8.1 \\
{\tt run104} & 12 & 36 & 0.09 & 0.16 & 1.94 & 0.84 & 1.11 & 1.36 & 4 & 65 & 33 & 8 & 0 & 3 & 42 & 23 & 4 & 0 & 1.2 & 9.4 \\
{\tt run105} & 12 & 15 & 0.18 & 0.31 & 0.67 & 0.58 & 1.15 & 0.89 & 1 & 24 & 44 & 6 & 4 & 0 & 17 & 31 & 3 & 2 & 1.4 & 5.9 \\
{\tt run106} & 12 & 23 & 0.18 & 0.30 & 1.14 & 0.79 & 1.13 & 1.03 & 1 & 41 & 65 & 7 & 1 & 0 & 27 & 41 & 6 & 1 & 1.3 & 7.3 \\
{\tt run107} & 12 & 30 & 0.18 & 0.29 & 1.07 & 0.94 & 1.27 & 1.02 & 2 & 61 & 74 & 8 & 2 & 1 & 43 & 62 & 2 & 2 & 1.3 & 7.1 \\
{\tt run108} & 12 & 38 & 0.18 & 0.27 & 0.93 & 3.95 & 1.40 & 0.50 & 3 & 70 & 66 & 7 & 5 & 2 & 43 & 52 & 5 & 0 & 1.7 & 10.1 \\
{\tt run109*} & 12 & 15 & 0.27 & 0.41 & 0.46 & 0.50 & 1.30 & 1.09 & 0 & 21 & 67 & 6 & 1 & 0 & 12 & 42 & 1 & 0 & 1.4 & 9.3 \\
{\tt run110} & 12 & 23 & 0.27 & 0.40 & 0.67 & 73.41 & 1.34 & 0.11 & 0 & 44 & 99 & 3 & 5 & 0 & 30 & 67 & 1 & 1 & 1.4 & 7.3 \\
{\tt run111} & 12 & 32 & 0.27 & 0.39 & 0.65 & 3.93 & 1.47 & 0.43 & 1 & 70 & 128 & 10 & 5 & 1 & 42 & 85 & 3 & 2 & 1.4 & 6.9 \\
{\tt run112} & 12 & 40 & 0.27 & 0.36 & 0.65 & 2.35 & 1.58 & 0.71 & 5 & 60 & 112 & 14 & 4 & 2 & 38 & 71 & 4 & 3 & 1.7 & 7.9 \\
{\tt run113*} & 12 & 13 & 0.05 & 0.10 & 1.93 & 1.36 & 0.80 & 0.54 & 1 & 16 & 9 & 0 & 0 & 1 & 7 & 6 & 0 & 0 & 1.6 & 5.7 \\
{\tt run114*} & 12 & 21 & 0.05 & 0.10 & 4.39 & 1.51 & 0.71 & 0.66 & 4 & 24 & 12 & 0 & 0 & 0 & 16 & 7 & 0 & 0 & 1.5 & 6.7 \\
{\tt run115*} & 12 & 28 & 0.04 & 0.09 & 7.53 & 5.29 & 0.69 & 0.47 & 3 & 34 & 17 & 0 & 0 & 2 & 16 & 11 & 0 & 0 & 1.6 & 4.7 \\
{\tt run116*} & 12 & 35 & 0.04 & 0.09 & 9.44 & 4.24 & 0.70 & 0.56 & 6 & 53 & 16 & 1 & 1 & 2 & 35 & 12 & 0 & 0 & 1.7 & 6.6 \\
{\tt run117*} & 12 & 14 & 0.09 & 0.17 & 0.97 & 0.34 & 1.00 & 1.07 & 2 & 18 & 10 & 4 & 2 & 0 & 11 & 6 & 0 & 0 & 1.5 & 4.1 \\
{\tt run118*} & 12 & 21 & 0.09 & 0.17 & 2.46 & 8.78 & 0.87 & 0.21 & 3 & 32 & 43 & 1 & 4 & 2 & 21 & 31 & 0 & 3 & 1.6 & 6.3 \\
{\tt run119*} & 12 & 29 & 0.09 & 0.16 & 4.32 & 1.06 & 0.84 & 1.00 & 4 & 50 & 41 & 1 & 1 & 0 & 25 & 28 & 0 & 0 & 1.5 & 5.9 \\
{\tt run120*} & 12 & 36 & 0.09 & 0.16 & 3.98 & 1.15 & 0.92 & 1.06 & 4 & 66 & 61 & 0 & 4 & 2 & 43 & 38 & 0 & 3 & 1.5 & 8.6 \\
{\tt run121*} & 12 & 14 & 0.18 & 0.30 & 0.70 & 2.15 & 1.15 & 0.39 & 0 & 23 & 37 & 5 & 0 & 0 & 13 & 24 & 2 & 0 & 1.6 & 7.3 \\
{\tt run122*} & 12 & 22 & 0.18 & 0.31 & 2.31 & 12.09 & 0.94 & 0.23 & 1 & 52 & 79 & 1 & 1 & 1 & 34 & 52 & 1 & 0 & 1.6 & 6.1 \\
{\tt run123*} & 12 & 30 & 0.18 & 0.29 & 3.39 & 3.21 & 0.90 & 0.49 & 3 & 53 & 102 & 0 & 2 & 1 & 32 & 60 & 0 & 1 & 1.4 & 7.6 \\
{\tt run124*} & 12 & 38 & 0.18 & 0.29 & 4.83 & 2.56 & 0.91 & 0.73 & 5 & 90 & 146 & 4 & 7 & 1 & 49 & 97 & 1 & 4 & 1.7 & 6.4 \\
{\tt run125*} & 12 & 15 & 0.27 & 0.42 & 0.95 & 11.82 & 1.04 & 0.15 & 1 & 22 & 47 & 2 & 0 & 0 & 10 & 26 & 0 & 0 & 1.4 & 6.6 \\
{\tt run126*} & 12 & 23 & 0.27 & 0.41 & 1.84 & 36.82 & 1.00 & 0.08 & 0 & 44 & 124 & 2 & 1 & 0 & 27 & 82 & 1 & 0 & 1.3 & 7.1 \\
{\tt run127} & 12 & 31 & 0.27 & 0.41 & 3.13 & 12.32 & 0.95 & 0.27 & 4 & 51 & 160 & 1 & 3 & 1 & 23 & 109 & 0 & 1 & 1.4 & 7.5 \\
{\tt run128} & 12 & 39 & 0.27 & 0.39 & 2.61 & 1.77 & 1.07 & 0.89 & 2 & 79 & 204 & 3 & 5 & 0 & 45 & 119 & 2 & 0 & 1.6 & 10.9 \\
\enddata

\tablecomments{$t_{\rm{cl}}$ denotes the cluster age in Gyr, 
$M$ denotes the total cluster mass at $t_{\rm{cl}}$ in $10^4\,\msun$, $f_b$ denotes the overall binary fraction 
in the cluster, $f_{b,c}$ denotes the binary fraction in the core, $\rho_c$ denotes the core density in 
$10^4\,\dens$, and $r_c$ denotes the core radius in pc. The corresponding ``observed'' values are given as $r_{c,\rm{obs}}$ and $\rho_{c,\rm{obs}}$. $N_{\rm{BSS}}$ and $N_{\rm{BSS},c}$ denote 
the numbers of BSSs in the whole cluster, and within $r_c$, respectively.  The time since formation 
for the BSSs is denoted by $t_{\rm{age}}$ in Gyr.  {\tt Coll}, {\tt MTB}, and {\tt SE} denote BSSs formed 
via stellar collisions, stable mass-transfer in a binary, and binary stellar evolution driven mergers, respectively.  
\sing, \bs, and \bb\ denote collisions via single-single, binary-single, and binary-binary encounters, respectively.  
Med, and Max denote the median $t_{\rm{age}}$ and the maximum $t_{\rm{age}}$ for each model. Models that are 
in the quasi-steady binary-burning stage \citep[equivalent to post core-collapsed;][]{2013MNRAS.429.2881C} are 
marked by ``*". }
\end{deluxetable}
\clearpage

\end{document}